\documentclass[preprint2]{aastex}
\newcommand{\etal}{{et al.}}

\newcommand{\lsun}{L$_{\odot}$}

\slugcomment{DRAFT v.2002.06.18}		

\begin{document}

\title{The Structure and Stellar Content 
of the Central Region of M33\altaffilmark{1}}

\author{Andrew W. Stephens\altaffilmark{2,3}}
\affil{The Ohio State University, Department of Astronomy}
\affil{140 West 18th Avenue, Columbus, OH  43210}

\author{Jay A. Frogel \altaffilmark{4}}
\affil{NASA Headquarters}
\affil{300 E Street SW, Washington, DC  20546}

\altaffiltext{1}{Based on observations obtained at the Gemini North Observatory.}
\altaffiltext{2}{Princeton-Catolica Prize Fellow}
\altaffiltext{3}{Current address:  Pontificia Universidad Cat\'{o}lica 
de Chile, Departamento de Astronom\'{\i}a y Astrof\'{\i}sica, 
Cassilla 306, Santiago 22, Chile; stephens@astro.puc.cl}
\altaffiltext{4}{Permanent address: The Ohio State University, 
Department of Astronomy, 140 West 18th Avenue, Columbus, OH  43210}

\begin{abstract}

Using Gemini QuickStart infrared observations of the central $22''$ of
M33, we analyze the stellar populations in this controversial region.
Based on the slope of the giant branch we estimate the mean metallicity
to be $-0.26 \pm 0.27$, and from the luminosities of the most luminous
stars, we estimate that there were two bursts of star formation $\sim 2$
and $\sim 0.5$ Gyr ago.  We show that the stellar luminosity function
not only has a different bright end cutoff, but also a significantly
different slope than that of the Galactic bulge, and suggest that this
difference is due to the young stellar component in M33.  We combine our
infrared Gemini data with optical HST-WFPC2 measurements revealing a CMD
populated with young, intermediate, and old age stellar populations.
Using surface brightness profiles from $0.1''$ to $18'$, we perform
simple decompositions and show that the data are best fit by a
three-component, core + bulge + disk model.  Finally, we find no
evidence for radial variations of the stellar populations in the inner
$3-10''$ of M33 based on a spatial analysis of the color-magnitude
diagrams and luminosity functions.

\end{abstract}


\section{Introduction} \label{sec:introduction}

With the advent of space-based telescopes, such as the Hubble Space
Telescope (HST), and large aperture ground-based telescopes with
adaptive optics (AO), such as Gemini and VLT, the number of galaxies
beyond the Milky Way (MW) and its dwarf companions for which detailed
studies can be made is gradually increasing.

Of the nearby galaxies, M33 (NGC 598, the Triangulum Nebula) is one of
the best for studying the stellar content of spiral galaxies.  Outside
of the MW, it is one of the closest and brightest spirals visible,
surpassed only by M31.  M31 is more luminous and slightly closer, but
it's higher inclination angle of 77\degr\ (compared to 56\degr\ for M33)
makes it more difficult to separate the contributions from different
stellar populations.  However, is M33 the prototype or the exception
when it comes to late-type spiral galaxies?  It contains a healthy
population of halo globular clusters \citep{Schommer1991}, yet it is not
certain whether there is a matching bulge.  This simple point is crucial
for understanding the role of a bulge in galaxy formation, and its
relationship to the halo and globular clusters.

As of 1991, the status of M33's bulge was ``controversial'' according to
a review of literature by \citet{vandenbergh1991}.  There have since
been many papers on M33, but the number of authors who find a bulge
seems balanced by an equal number who do not. \citet{Bothun1992} argued
against a traditional bulge, based on the lack of a power-law
contribution to his 12\micron\ surface brightness measurements.
Although his $B$-band observations do show an excess of light inside
$3'$, he finds no satisfactory $r^{1/4}$ fit.  Based on infrared (IR)
observations of the central $2.5' \times 8'$ which show a clustering of
stars around the nucleus, \citet{Minniti1993, Minniti1994} take the
opposite side.  They find a de Vaucouleurs profile fits all but the
inner $1'$ of their surface photometry.  \citet{Regan1994} are also in
favor of a bulge in M33.  Their infrared surface brightness measurements
of the central $15' \times 30'$ seem reasonably well fit by an
exponential disk plus $r^{1/4}$ profile.  In contrast,
\citet{McLean1996} cite an unchanging IR luminosity function from $45''$
to $1.5'$ as evidence against a significant bulge population beyond
$45''$.  Thus, a decade after van den Bergh's review, the controversy
remains unresolved.

In this paper we use IR Gemini-North observations with QUIRC/Hokupa'a to
study the stellar populations in the inner regions of M33.  In section
\ref{sec:observations} we describe our observations and observing
procedure.  Section \ref{sec:data_reduction} details the data reduction,
photometric techniques, and calibration.  We give the azimuthally
averaged surface brightness profile in Section \ref{sec:decomposition}
and, after combining our observations with those of \citet{Regan1994},
present simple two- and three- component model decompositions.  We
describe artificial star tests performed to understand completeness and
observational effects as a function of a star's position and luminosity
in Section \ref{sec:artstartests}.  We use the color-magnitude-diagram
(CMD) to estimate the metallicities and ages of the stars in our field
in Section \ref{sec:cmds}.  We present the stellar luminosity function
(LF) in Section \ref{sec:lfs}, and compare it to the LF measured in the
Galactic bulge.  In section \ref{sec:comparison} we compare our
observations to the recent work in the infrared by \citet{Davidge2000a}
with the CFHT AO system, and combine our data with optical HST-WFPC2
measurements made by \citet{Mighell2002}. We look for radial variations
in the stellar properties in Section \ref{sec:radial_variations}. In
Section \ref{sec:blending} we perform a theoretical analysis of blending
on our own and previous observations following the procedures of
\citet{Renzini1998}.  Finally we give a summary of our conclusions in
Section \ref{sec:conclusions}.

\section{Observations} \label{sec:observations}

The observations upon which this paper is based were taken as part of the
Gemini North QuickStart Service Observing Program using Hokupa'a \&
QUIRC.  Hokupa'a \citep{Graves1998} is a natural guide star, 36-element
curvature-sensing adaptive optics system built by the University of
Hawaii.  QUIRC is a near-infrared imager on loan from the University of
Hawaii and mounted at the exit focus of Hokupa'a.  QUIRC has a $1024
\times 1024$ HAWAII HgCdTe array with a plate scale of $0.02''$
pixel$^{-1}$, giving a $\sim 20.5''$ field of view.  The array is linear
to $\sim 40,000$ ADU, and saturates at $50,000$ ADU, with a gain of 1.85
$e^-$/ADU.

We observed the central regions of M33 ($\alpha=$01:33:50.9, 
$\delta=+$30:39:37, J2000) through three filters, $J$, $H$, and $K'$
\citep{Wainscoat1992}.  We used the nucleus of M33 as the AO wavefront
reference source.  Observations were carried out over three separate
nights spanning three months.  Table \ref{tab:observations} lists the
number of successful frames which were obtained in each band on each
night.  Exposure times were 180 seconds except on October 25 when
observations were made without a shutter; on that night the exposure
times were approximately 10.8 seconds longer due to the detector readout
time.

\begin{figure*}[ht]
\epsscale{2.1}
\plotone{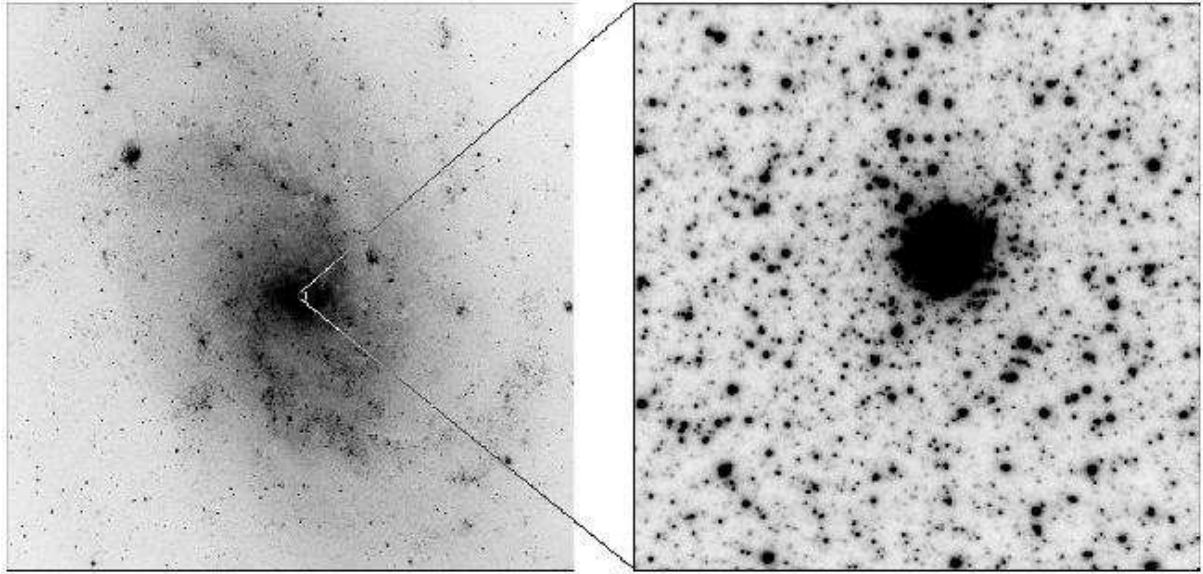} 
\figcaption{
Relative size and location of our infrared image.  The left image is a
$30'$ red Digitized Sky Survey image centered on M33, and on the right
is our $\sim 22''$ $H$-band Gemini-North image.  North is up and east is
to the left.  The faintest stars visible in the infrared image have $H
\sim 20.5$.
\label{fig:field_locator}}
\end{figure*}

The relative size, location, and orientation of our field is illustrated
in Figure \ref{fig:field_locator}.  On the left of this Figure is a
$30'$ Digitized Sky Survey\footnote{The Digitized Sky Surveys were
produced at the Space Telescope Science Institute under U.S. Government
grant NAG W-2166.}  image centered on M33.  The box drawn in the center
of the Galaxy represents the placement of our infrared Gemini-North
image.  The infrared image shown on the right is our combined $H$-band
image, and is $\sim 22''$ across.

We obtained sky observations for background subtraction in a blank field
$\sim 45'$ SE of the nucleus ($\alpha=$01:34:57, $\delta=+$29:57:20,
J2000).  The typical observing pattern was to take two sky dithers, four
dithers on the nucleus, two skys, etc.  With this technique the sky
frame is constructed from the combination of the pairs of sky frames
before and after each observation, and hence each two-dither sky set is
used for both the preceding and following galaxy observations.  The sky
sets used $3.2''$ dithers, while the galaxy sets used $1.3''$ dithers.

UKIRT photometric standard stars were observed each night, although
possibly not quite as frequently as the authors would have liked.  The
September night had one standard before and one after our observations;
the October night had two before; and the December night had two before
and one after.  These standards were typically observed with 5 second
exposure times, implementing a five position dither pattern moving the
star in $10''$ steps.

\begin{deluxetable}{cccc}
\tablewidth{0pt}
\tabletypesize{\footnotesize}
\tablecaption{Dates \& Number of Observations}
\tablehead{
\colhead{Date}		&
\colhead{N($J$)}	&
\colhead{N($H$)}	&
\colhead{N($K'$)}	}
\startdata
2000-09-27 &  8  &  0  & 12  \\
2000-10-25 &  8  &  0  &  8  \\
2000-12-22 &  0  & 19  &  0  \\
\enddata
\label{tab:observations}
\end{deluxetable}

\begin{figure}[ht]
\epsscale{1}
\plotone{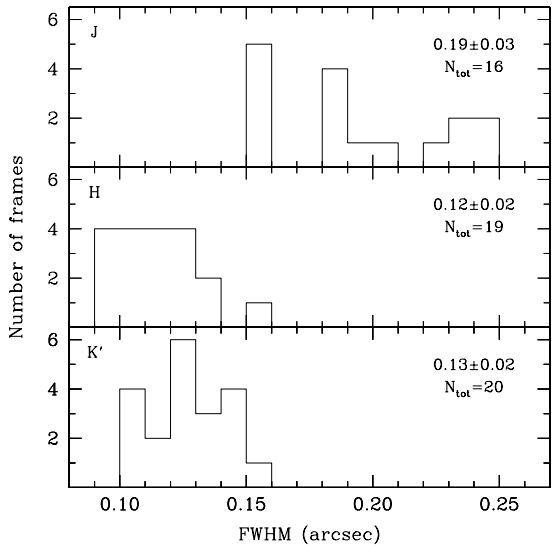} 
\figcaption{
The observed distribution of image quality for all observations.  The
upper panel shows the measured FWHM in arcseconds for the $J$-band
frames.  Similar histograms for the $H$ and $K'$ bands are displayed in
in the middle and lower panels respectively.  The average FWHM and the
total number of frames measured in each band is given in the upper right
corner of each panel.
\label{fig:fwhm}}
\end{figure}

Image quality was extremely good.  We used the IRAF\footnote{IRAF is
distributed by the National Optical Astronomy Observatories, which are
operated by AURA, Inc., under cooperative agreement with the NSF.} {\sc
gemseeing} procedure to estimate the full-width at half-maximum (FWHM)
on all 55 images.  The results shown in Figure \ref{fig:fwhm} are the
average FWHMs measured on the same $\sim 50$ hand-selected stars on all
frames.  The $K'$-band images have a mean seeing of $0.13'' \pm 0.02''$,
$H$-band obtained $0.12'' \pm 0.02''$, while the $J$-band images are
slightly worse and more variable with a mean seeing of $0.19'' \pm
0.03''$.  The superior image quality at longer wavelengths is a
consequence of the increased coherence length of the atmosphere
($\propto \lambda^{6/5}$).  The image size does not vary significantly
across the field, although the PSF morphology has a systematic
dependence on the azimuthal angle around the nucleus of M33, which was
used as the AO reference.  For comparison, the diffraction limits of an
8m telescope are $0.04''$, $0.05''$, and $0.07''$ at $J$,$H$, and $K'$
respectively.

The FWHM, however, is not the best measurement of image quality since
most AO images consist of a tight core superimposed on much broader
emission.  We therefore also recorded the {\sc gemseeing} task's
estimates of various encircled energy diameters (EED).  We found that
the 50\% EED is approximately twice the FWHM, while the average 95\% EEDs
are $0.82''$, $0.60''$, and $0.65''$ at $J$, $H$, and $K'$ respectively.

\section{Data Reduction \& Photometry} \label{sec:data_reduction}

We reduced our data using the Gemini package of tools in IRAF.
The procedure involves division by a normalized lamps on -- lamps off
dome flat field, and subtraction of a sky frame constructed from sky
observations before and after each galaxy observation.  A constant equal
to the median of the sky frame was then added back to the image to
restore the original sky level and maintain the noise characteristics of
the frame.

Photometry was performed with Peter Stetson's suite of software.  Object
detection was done on a combined image made up of all the dithers of all
the bands.  We then used {\sc daophot} \citep{Stetson1987} to determine
point spread functions (PSFs) for each dither of each band using
isolated stars.  We used a quadratically variable model PSF with a 25
pixel ($0.5''$) radius; an alternate run using a larger 50 pixel ($1''$)
radius PSF, yielded little difference in the photometry, while taking a
significantly larger amount of processing time.  Instrumental magnitudes
were measured using the {\sc allframe} PSF fitting routine
\citep{Stetson1994} which simultaneously fits PSFs to all stars on all
dithers.  {\sc daogrow} \citep{Stetson1990} was then used to determine
the best magnitude in a $0.5''$ radius aperture.

Our first attempt at calibrating the data used the photometric standards
which were observed for us.  We measured each standard using simple
aperture photometry with a $0.5''$ radius aperture, and a sky annulus
from $2''$ to $3''$.  Due to the small number of observed standards, we
combined the measurements from all three nights.  This gave us three $J$
standards and four $K'$ standards for the September and October nights,
and three $H$ standards for the December night.  Unfortunately, there
are no published $K'$ measurements for UKIRT standards.  We therefore
estimated $K'$ for each of the observed standards using the ($H-K$)
color and the transformation of \citet{Wainscoat1992}. Using the median
combination of these measurements, we estimated our photometric zero
point with uncertainties of 0.02 magnitudes in $H$ and $K'$, and 0.04
magnitudes in $J$.  We then converted our science $K'$ measurements to
$K$ based on each star's $(H-K')$ color using a transformation derived
from the data published by \citet{Wainscoat1992}.  The last step was to
convert our measurements to the CIT/CTIO photometric system using the
transformation equations of \citet{Hawarden2001}.

This initial calibration seemed to work acceptably for the $J$ and $H$
bands, but {\em not} for $K$.  Evidence of the failure was in the
disagreement ($\sim 0.3$ magnitudes) with previously published data for
the stars in the inner region of M33.  Specifically, with the $K$-band
measurements of bright stars measured by \citet{Davidge2000a}, and an
equal offset in the $K$-band surface brightness measurements published
by \citet{Regan1994}.  One possible cause of the problem is an
inadequate transformation between $K$ and $K'$.  The
\citet{Wainscoat1992} transformation was published in a paper describing
the original $K'$ filter.  Yet even a cursory comparison of the $K'$
filter transmission curve published for
QUIRC\footnote{\url{www.ifa.hawaii.edu/instrumentation/quirc/quirc.html}}
and that in the \citet{Wainscoat1992} article reveals that they are very
different.  As an example, the original filter had a peak transmission
of $\sim 80$\%, while the QUIRC $K'$ filter peaks over 95\%.

To overcome this problem we had to find alternate ``standards'',
preferably ones which had been observed before.  For this purpose, we
matched observations of 31 stars in common with the photometry published
by \citet{Davidge2000a}. We examined the difference between his $K$ band
observations (UKIRT photometric system) and our instrumental $K'$ for
any dependence on $(J-K)$ or $(H-K)$ color, but found no significant
trend.  We therefore adopted the mean difference as our photometric zero
point and conversion to the $K$-band.  The uncertainty in this zero point
is 0.02 magnitudes.  A more detailed discussion of the comparison
between these two datasets can be found in Section \ref{sec:davidge}.

With the measurements finally calibrated we could construct our final
photometry list.  Since our transformation to the CIT/CTIO system
requires both $(J-K)$ and $(H-K)$ color information, we had already
selected out only stars which have measurements in all three bands. In
addition, each star had to have been detected on at least five dithers
in each band, and the final error had to be less than 0.25 magnitudes.
This gave us a list of 3716 objects over the entire frame.  After
ignoring objects detected within $3''$ of the center of M33 because of
severe crowding and incompleteness (see Section \ref{sec:completeness}),
we were left with a final photometry list of 3308 measurements, covering
an area of 416 square arcseconds.

When converting to absolute or bolometric magnitudes, we will use the
distance and reddening to M33 as determined by \citet{Freedman1991} from
observations of Cepheid variables.  They find a true distance modulus of
$(m-M)_0=24.64$ (850 kpc, $1''=4.1$ pc), assuming 18.5 for the LMC.
They also measure the M33 reddening to be $E(B-V)=0.1$, assuming 0.1 for
the LMC, and a ratio of total-to-selective extinction of $R_V=3.1$.
These values translate to an infrared extinction and reddening of
$A_K=0.03$ and $E(J-K)=0.05$.  We note, however, that
\citet{Sarajedini2000} have recently calculated a slightly farther
distance of 930 kpc, $(m-M)_0=24.84 \pm 0.16$, using RR Lyrae's in two
halo clusters.


\section{Surface Brightness Decomposition} \label{sec:decomposition}

In order to obtain accurate surface brightness measurements, we started
with our reduced frames, and subtracted off the constant sky value which
had been added previously to maintain each frame's noise characteristics.  
We then converted each frame to count rates, since some of the frames had 
slightly different exposure times.  The resulting set of images in each 
band were averaged together to create the final images.

\begin{figure*}
\epsscale{2.3}
\plotone{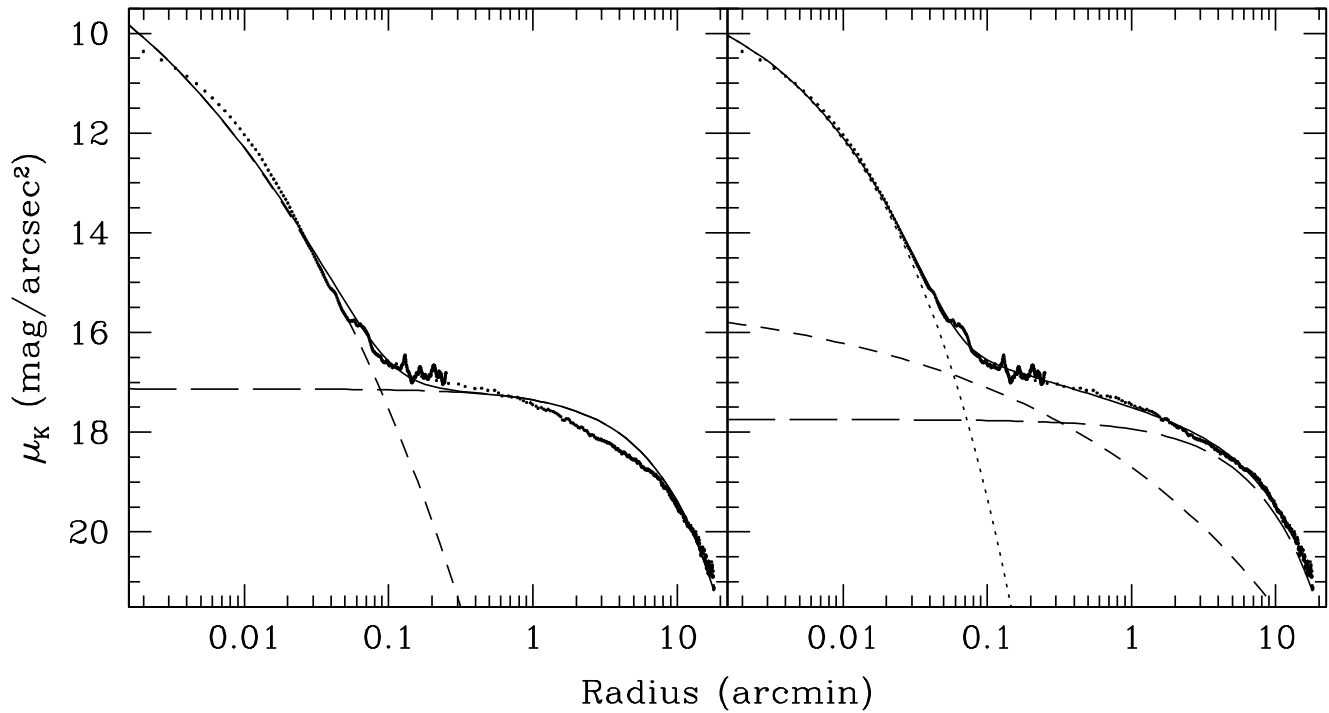} 
\figcaption{
The azimuthally averaged $K$-band surface brightness profile ($0.12'' <
r < 15''$) combined with measurements from \citet{Regan1994} ($7'' < r <
18'$).  The left panel shows the results of a simple 2-component
exponential disk (long dashed line) + $r^{1/4}$ bulge (short dashed
line) decomposition.  The right panel adds a Sersic core (dotted line)
as a 3rd component to the decomposition.
\label{fig:decomposition}}
\end{figure*}

We used the {\sc xvista} {\sc annulus} routine to compute azimuthally
averaged radial surface brightness profiles.  Assuming an inclination of
56\degr \citep{Zaritsky1989} and a position angle of 23\degr\ for M33,
the routine defines elliptical annuli to be used as the averaging paths,
in effect calculating the deprojected surface brightness profile
assuming a circular disk.  With 2 pixel wide annuli ($0.04''$), we were
able to calculate the surface brightness profile out to $15''$.  We then
combined our surface brightness measurements with those of
\citet{Regan1994} which span from $7''$ to $18'$.  The resulting
$K$-band profile is shown in Figure \ref{fig:decomposition}.  There are
three \citet{Regan1994} points in the region of overlap ($7''-15''$),
and in this region the average difference between the two datasets is
less than 0.01 magnitudes arcsecond$^{-2}$.  Note that since the surface
brightness information is extracted directly from pixel values, the
profile is unaffected by image blending, even in the most crowded regions.

Using this combined surface brightness profile ($0.12'' < r < 18'$), we
have performed a simple bulge/disk decomposition.  We assume an
exponential disk (eqn. \ref{eqn:disk}) with a central intensity $I_0$
and a disk scale length $R_d$, and a de Vaucouleurs $r^{1/4}$ bulge
(eqn. \ref{eqn:bulge}) with an effective radius $R_e$ and an intensity
$I_e$ at $R_e$.  The factor $a = \sqrt{(1 + \cos^2 i)/2}$ is a
correction factor for deprojecting a spherical bulge.

The best fitting 2-component model is illustrated in the left panel of
Figure \ref{fig:decomposition} with a solid line.  The disk has a
central surface brightness of 17.16 magnitudes arcsec$^{-2}$, and a scale
length $R_d = 4.87'$.  The bulge is very compact, with an effective
radius $R_e = 1.1''$, where the surface brightness is 13.87 mag
arcsec$^{-2}$.

The results of the fit are not very encouraging for this oversimplistic
disk + bulge model.  It is very clear that M33 has a strong compact
spheroid component which becomes dominant over the disk at $r \lesssim
5''$.  However, the best fit parameters are neither a good fit to the
inner bulge, nor to the disk between $1' - 7'$.

\begin{equation}
I_{disk}(r) = I_0  \exp(-r/R_d)
\label{eqn:disk}
\end{equation}
\begin{equation}
I_{bulge}(r) = I_e  \exp \{ -7.67 [ (ar/R_e)^{1/4} - 1] \}
\label{eqn:bulge}
\end{equation}
\begin{equation}
I_{core}(r) = I_s \exp \{ -b_n  [ (ar/R_s)^{1/n} - 1] \}
\label{eqn:core}
\end{equation}

A simple alternative is to add a third component to our model.  We thus
try an exponential disk + $r^{1/4}$ spheroid + Sersic core
\citep{Sersic1968}.  The Sersic function (eqn. \ref{eqn:core}) is
parameterized with an effective radius $R_s$, an intensity $I_s$ at
$R_s$, and the shape index $n$ ($b_n \approx 1.9992 n - 0.3271$).  The
variable power-law index $n$ gives the Sersic function an extra degree
of freedom over the more constrained exponential or $r^{1/4}$ profiles.

The results of this three component model are shown on the right side of
Figure \ref{fig:decomposition} and the parameter values are listed in
Table \ref{tab:decomposition}.  Not surprisingly, this yields a
significantly better fit than for the two component model.  The central
peak in surface brightness is well described by the Sersic function, the
disk is $\sim 0.5$ magnitude fainter, and the spheroid explains the excess
luminosity seen inside $1'$.  Note that we list only the parameters
derived from the $H$- and $K$-band images.  The $J$-band image quality
was inadequate to accurately model the very narrow core component.

\begin{deluxetable}{lcc}
\tablewidth{0pt}
\tablecaption{Three Component Model Parameters}
\tabletypesize{\footnotesize}
\tablehead{
\colhead{Parameter}	& 
\colhead{$H$-band}	&
\colhead{$K$-band}	}
\startdata
Disk scale length, $R_d$	&  $5.58'$  &  $5.67'$	\\
Disk central SB, $I_0$		&  18.13    &  17.75	\\
Spheroid effective radius, $R_e$& $39.99'$  & $22.36'$	\\
Spheroid SB @ $R_e$, $I_e$	&  24.08    &  23.40	\\
Core effective radius, $R_s$	&  $0.61''$ &  $0.68''$	\\
Core SB @ $R_s$, $I_s$		&  13.03    &  12.73	\\
Core shape index, $n$		&   2.21    &   2.07	\\
\enddata
\tablecomments{Surface brightnesses are in mag/arcsec$^2$.}
\label{tab:decomposition}
\end{deluxetable}

These disk and spheroid parameters are similar to those derived by
\citet{Regan1994} based on their $r>7''$ data alone.  They found a disk
scale length of $6.1' \pm 0.3'$ at $H$ and $5.8' \pm 0.4'$ at $K$, with
a central surface brightnesses of $H=18.21$ and $K=17.80$
mag/arcsec$^{2}$.  However, since they were only using a two-component
model (and excluding the inner $6''$), they derived a brighter, more
compact spheroid with $R_e = 8'$, $I_e(H)=22.84$, and $I_e(K)=22.63$
magnitudes arcsecond$^{-2}$.  Their more compact spheroid is also in
agreement with the $8.5'$ spheroid effective radius found by
\citet{Minniti1993}.

\citet{Kent1987} studied the surface brightness profile of M33 in the 
visible, and found a larger exponential disk scale length of $9.6'$.  He
also realized that the profile was too high for just an exponential disk
inside $3'$, but he dismissed it due to the presence of visible spiral
structure, concluding that the disk is non-exponential in the center.

If our three component model is taken literally, it implies that
virtually our entire field is spheroid dominated.  At $3.6''$ the spheroid and
central core have equal surface brightness, which is 0.7 magnitudes
brighter than that of the disk.  At the edges of our field, $r \sim
15''$, the disk and spheroid surface brightness become nearly equal.  If
this is correct, and the spheroid and disk are composed of different
stellar populations, we may be able to detect a very small change in the
composite population as a function of radius (see Section
\ref{sec:radial_variations}).

We have also estimated the size of the core from our combined $K'$
image.  This image has better image quality than the $J$ band, and is
not saturated like the $H$ images.  On this image the nucleus has a FWHM
of $0.30''$, compared to the stars which have a width of $0.13''$.  Thus
we estimate that the nucleus has a width in $K'$ of less than $0.27''$.
We give this value as an upper limit since we do not attempt to correct
for the nonlinearity of the array in the very core ($r < 0.1''$) where
the count rate is too high for the exposure time.  This value seems in
line with the increasing core size with wavelength measured by
\citet{Gordon1999} using HST.  They find a core FWHM of $0.12''$ with the
F300W filter, $0.20''$ with the F555W filter, and $0.25''$ with the
F1042 filter.

Although we could not measure the nuclear luminosity because the central
pixels of our images were saturated, we can make an estimate based on
our surface brightness fits.  Integrating the ``core'' component from
the best-fitting three component model, we find a total luminosity of
$M_H = -14.03$ and $M_K = -14.54$.  Assuming $M_{K\odot} = +3.33$, this
corresponds to $L_K = 1.4 \times 10^7 L_{\odot}$.

%
%


\section{Artificial Star Tests} \label{sec:artstartests}

In order to understand the completeness of our observations and estimate
the importance of blending, we have performed a series of tests using
artificial data.  The traditional completeness tests inject artificial
stars into the observed frames; the efficiency and accuracy of their
recovery give information about the completeness of the observations
\citep{Stetson1988}. This method is straightforward and easy to use, but
is limited to primarily uncrowded fields.  The more complex artificial
field technique creates an entire frame containing millions of
artificial stars to match the observations
\citep{Stephens2001}.  This method is especially useful in very crowded
fields, providing important constraints on the underlying stellar
population.

\begin{figure}[ht]
\epsscale{1}
\plotone{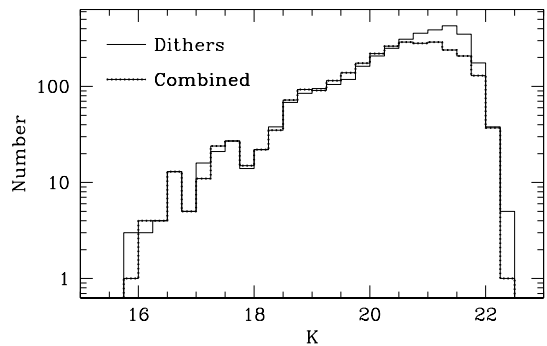} 
\figcaption{
Comparison of luminosity functions obtained from photometry of all 55
individual, uncombined dithers (solid), and photometry from the final
combined images (beaded).  Both LFs only include stars measured farther
than $3''$ from the center of M33.  The LFs are virtually identical
except at the faint end, where the combined frame photometry turns over
at $K \sim 20.5$, while the photometry performed on individual dithers
does not turn over until $K \sim 21.5$.
\label{fig:dither_vs_combined_lfs}}
\end{figure}

However, since photometry of the entire set of 55 $1024 \times 1024$
pixel frames can be quite time consuming, taking many days to complete,
our artificial star tests were instead run on the final combined $JHK$
images.  To validate this shortcut, we have compared photometry obtained
from the combined images with that from the simultaneous measurements of
the 55 individual dithers.  Tests show that these two techniques are
nearly identical, except for the faintest stars, where the photometry
performed on the individual frames tends to be more complete.  This can
be seen in Figure \ref{fig:dither_vs_combined_lfs}, which shows
luminosity functions derived from each of the two methods.  The LF
obtained from the individual frame photometry (solid line) rolls over
approximately one magnitude fainter than the LF from the combined frames
(beaded line), even though they both go to about the same depth.

\subsection{Traditional Completeness Tests} \label{sec:completeness}

Starting with the three ($JHK'$) final combined images, we used the {\sc
daophot} {\sc addstar} routine \citep{Stetson1987} to insert 361
artificial stars into each image.  These stars were arranged in a grid
of $19 \times 19$, with a spacing of $2 \times R_{PSF} + 2 = 52$ pixels
($\sim 1''$), to avoid self-crowding.  Each star included random Poisson
noise, and random sub-pixel coordinate offsets from the aforementioned
grid.  All stars were input with the same magnitude starting with
$K_i=17$, and a color based on the mean colors observed in the real
frame.  We then repeated this process, with the grid of input stars
shifted by 26 pixels ($\sim 0.5''$).  After four trials we had added a
total of 1444 artificial stars, each sampling different positions in our
frame.

After the addition of the artificial stars, we analyzed these frames in
the same manner as our original frames.  Stars were detected with {\sc
daofind} and measured on each of the four sets of $JHK'$ images
simultaneously with {\sc allframe}.  Once the final photometry list was
completed, we extracted the colors and magnitudes recovered for the
input artificial stars using {\sc daomaster} to match recovered
coordinates with input coordinates, requiring less than a 2 pixel
($0.04''$) difference in position.

\begin{figure}[ht]
\epsscale{1}
\plotone{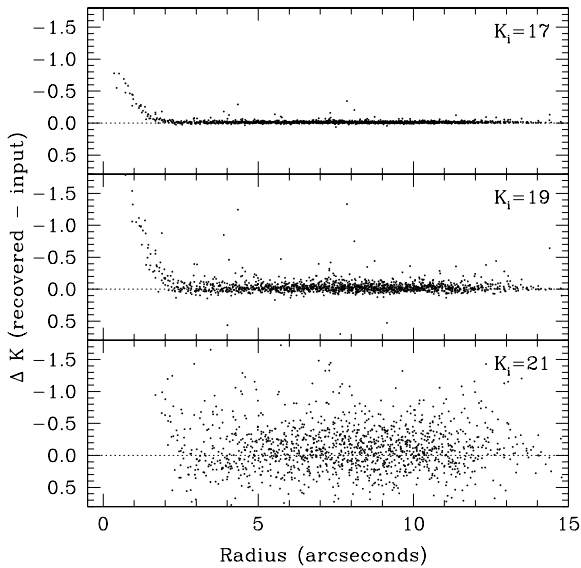} 
\figcaption{
Difference between recovered and input artificial star magnitudes as a
function of distance from the center of M33.  The upper panel shows
stars with an input $K$-band magnitude of 17, the middle shows stars
input at $K_i=19$, and the lower panel is for $K_i=21$.
\label{fig:deltamag}}
\end{figure}

We repeated this 4-trial procedure for six different integer input
magnitudes ($K_i=17-22$).  We have thus accumulated statistics on a
total of 8664 artificial stars.  Figure \ref{fig:deltamag} shows the
difference between the recovered and input magnitudes as a function of
distance from the center of M33 for three of the six input magnitudes
($K_i=\{17,19,21\}$).  As expected the brightest stars are recovered
very accurately, with only a few outliers caused by blends with other
bright stars on the frame.  Note that an artificial star has to fall
almost exactly on top of another star to be brightened but not lost
since its position cannot be perturbed by more than $0.04''$ to be
considered the same input star.  Near the center of M33 ($r\lesssim
3''$), where the density of stars is very high, Figure
\ref{fig:deltamag} shows that almost all of the artificial stars are
measured significantly brighter than their input magnitude.

\begin{figure}[ht]
\epsscale{1}
\plotone{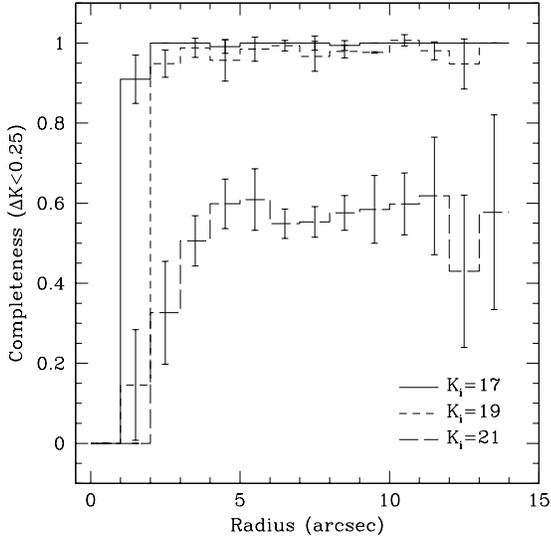} 
\figcaption{
Completeness as a function of distance from the center of M33 for three
different input magnitudes: $K_i=17$ (solid), $K_i=19$ (short dashed), and
$K_i=21$ (long dashed).  We have required that the absolute difference
between the recovered and input magnitudes be no greater than 0.25
magnitudes.
\label{fig:comp_vs_rad}}
\end{figure}

Since the surface brightness varies across the field, our photometric
completeness will also be a function of position on the frame.  In Figure
\ref{fig:comp_vs_rad} we plot the completeness as a function of radius
for three of the six input magnitudes.  To estimate the completeness, we
have counted up the number of recovered stars in radial bins around the
center of M33, requiring that the difference between input and recovered
magnitudes be less than 0.25.  We then divide by the number of
artificial stars input into these same bins.  The results are averaged
over the four trials, and the errorbars show the one sigma dispersion in
the range of values obtained over the four trials.

This plot shows several important things about the completeness across
our frame.  First, the completeness drops off very quickly near the
center of M33.  The completeness of the brighter stars drops from almost
100\% to near zero across a range of only $1''$.  The fainter stars do
not exhibit quite as precipitous a drop, but still make the transition
from maximum to minimum values quite quickly; albeit at a larger central
distance.  Second, the completeness away from the nucleus is relatively
constant.  Thus it is safe to simply throw away measurements inside some
critical radius, where the completeness is low (and the effects of
blending are high), and afterwards not worry about spatially varying
completeness.  Finally, we can measure no stars accurately inside $1''$.
While some stars are recovered in this region (see Figure
\ref{fig:deltamag}) their measured brightnesses are all discrepant by
more than 0.25 magnitudes.

\begin{figure}[ht]
\epsscale{1}
\plotone{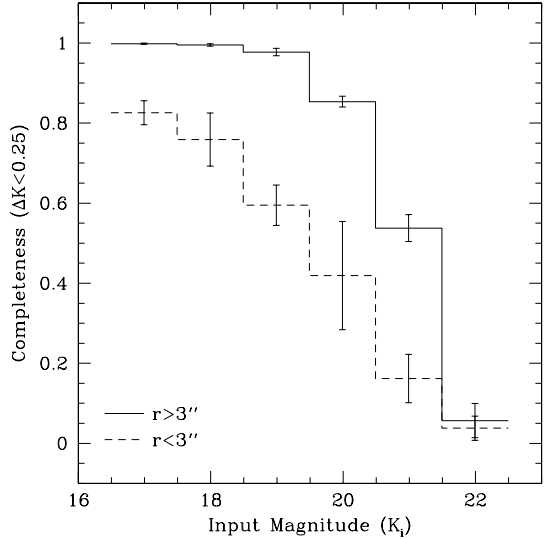} 
\figcaption{
Completeness as a function of the input $K$-band magnitude plotted for
regions farther than $3''$ (solid) and closer than $3''$ (dashed) from
the center of M33.  To be counted as recovered, a star had to be
measured to within $0.04''$ and 0.25 magnitudes of its input position
and magnitude.
\label{fig:comp_vs_mag}}
\end{figure}

Figure \ref{fig:comp_vs_mag} shows the photometric completeness as a
function of the brightness of the input artificial stars.  Here again we
require that the recovered magnitude be no more than 0.25 magnitudes
different than the input magnitude.  Since Figure \ref{fig:comp_vs_rad}
shows that the negative effects of the core of M33 are constrained to
the inner $\sim 3''$, we split this figure into two components.  The
solid line shows the completeness measured for all stars outside $3''$,
which are again lower limits on the completeness since measured on the
combined frames, while the science photometry was performed on all 55
individual frames simultaneously.  The dashed line is for all objects
measured closer than $3''$ to the center of M33, and shows that the
photometry inside $3''$ is severely degraded at all magnitudes.

\subsection{Artificial Fields}

We have created a completely artificial field to match our M33
observations.  We processed and measured this field in exactly the same
manner as the real observations.  Since we know both the measured and
true magnitude and color of every star in the artificial field, we can
quantify our errors and try to estimate the true properties of the
observed stellar population being modeled, free from observational
effects.

Starting with blank frames with the appropriate amount of Poisson noise,
we add 8 million stars using spatially variable PSFs modeled from the
real data.  The stars follow the observed M33 radial distribution,
specifically the sum of the three component $K$-band decomposition of \S
\ref{sec:decomposition}.

\begin{figure}[ht]
\epsscale{1}
\plotone{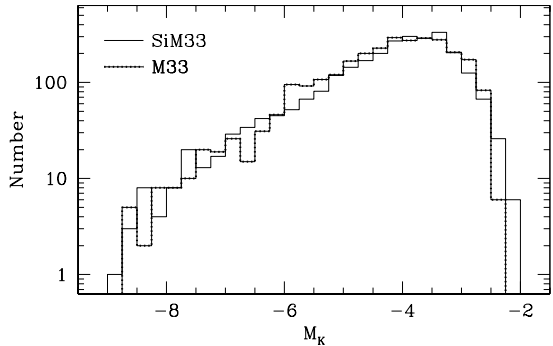} 
\figcaption{
Luminosity functions recovered from the simulated M33 frame (solid line)
and the real combined frames (beaded line).  In total, we measured 2802
objects in the real (combined) frame and 2677 in the simulation ($r>3''$).
\label{fig:simlfs}}
\end{figure}

The input stellar population is chosen to match the colors and
luminosity function observed in our M33 field.  The input colors are the
mean colors observed for stars with $r>3''$, and calculated at 0.5
magnitude intervals.  The input luminosity function follows a power law
distribution with a slope $d(log N)/dM_K = 0.37$ (as is measured in \S
\ref{sec:lfs}), extending from $-8.8<M_K<5$.  We add stars until the
recovered LF of the simulation approximately matches the recovered LF of
the real observations.  In this case we added 8 million stars, which
gives us slightly fewer recovered stars than in the real frame, 2677
artificial vs 2802 real, however as Figure \ref{fig:simlfs} shows, we
get a very good match over most of the range of luminosities.

The artificial frames were then processed and measured in the same
manner as the real data, namely finding stars on a combined image with
{\sc daofind}, modeling the PSF from isolated stars, and simultaneous
PSF-fitting photometry with {\sc allframe}.

\begin{figure}[ht]
\epsscale{1}
\plotone{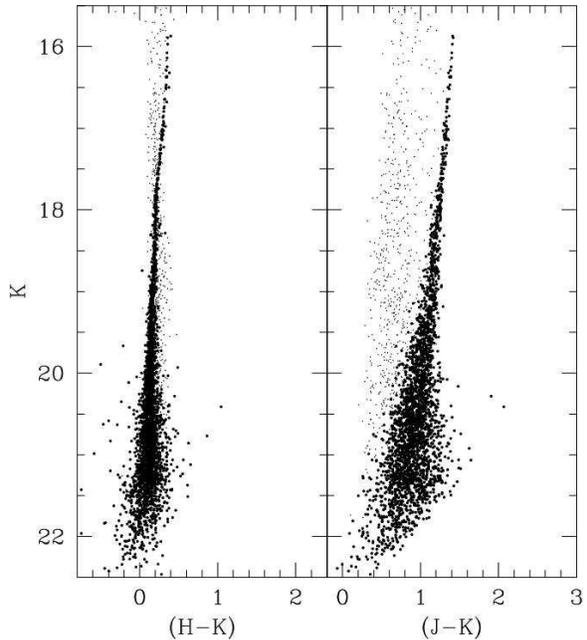} 
\figcaption{
The CMDs measured from the simulated field.  Half size points indicated
objects measured with $r<3''$.
\label{fig:simcmd}}
\end{figure}

The resulting $M_K$,$(J-K)$ and $M_K$,$(J-H)$ CMDs are shown in Figure
\ref{fig:simcmd}.  Here we have included all objects measured on the 
frame to show how well the simulation reproduces the observations,
including blended objects located within $r=3''$ (half-size points).  We
will discuss the implications of these simulations when we attempt to
quantify the effects of crowding on our photometry.

\section{Color -- Magnitude Diagram} \label{sec:cmds}

\begin{figure*}[ht]
\epsscale{1.5}
\plotone{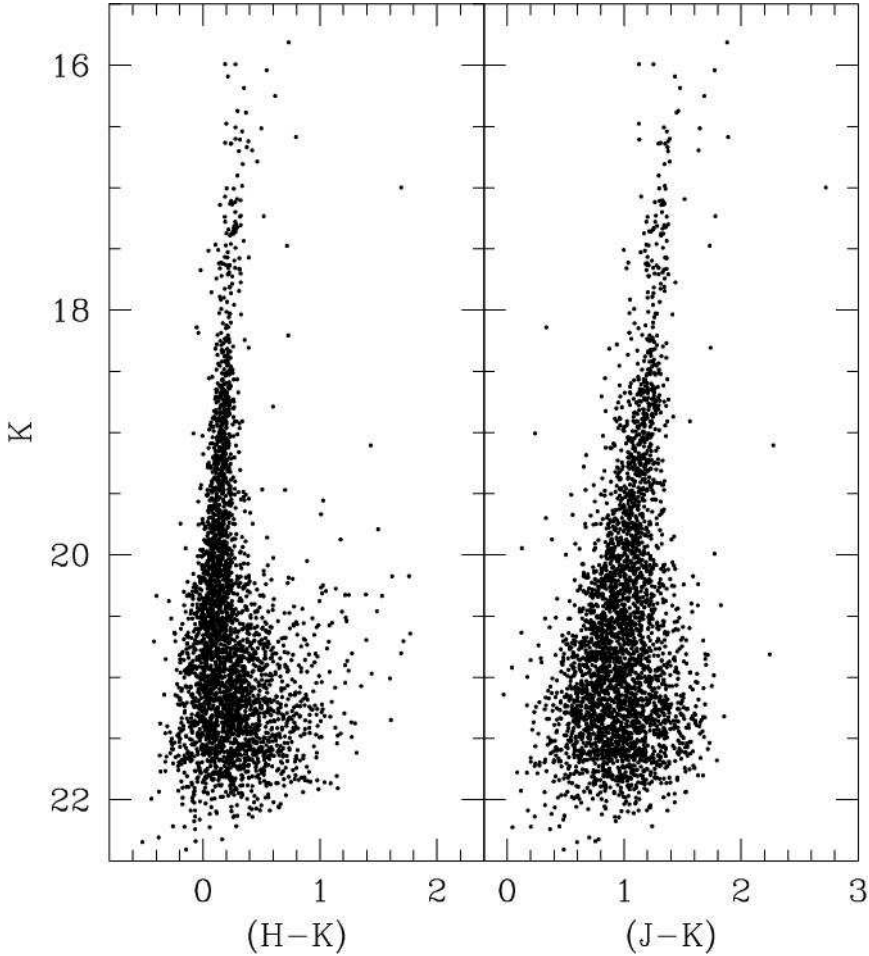} 
\figcaption{
Color magnitude diagrams for our M33 field ($r>3''$) obtained from
simultaneous photometry on the individual dithers.
\label{fig:cmd}}
\end{figure*}

The CMD of our M33 field is shown in Figure \ref{fig:cmd}.  Here we have
omitted objects measured within $3''$ of the nucleus which are mostly
blends of fainter, bluer stars as indicated by their characteristic
blueward and upward shift from the main giant branch locus (see Figure
\ref{fig:simcmd}).  We have shown in Section \ref{sec:completeness} that
nearly all objects detected close to the nucleus are blended, and that
$3''$ is approximately the boundary outside of which accurate photometry
can be achieved.  We thus take $3''$ as the dividing radius between
``good'' and ``bad'' photometry.  The resulting CMD for $r>3''$ is
relatively clean; we interpret the gap at $K \sim 18.2$ as the tip of
the red giant branch (RGB), and stars extending to $K \sim 16$ as
the asymptotic giant branch (AGB).  We compare our CMD with measurements
by others in Section \ref{sec:comparison}.

\subsection{Metallicity} \label{sec:metallicity}
The slope of the RGB is very sensitive to the metallicity of the
population.  \citet{Kuchinski1995a} and \citet{Kuchinski1995b} have used
this fact to develop a metallicity indicator using the slope
$(\Delta(J-K)/\Delta M_K)$ of the infrared RGBs of the upper five
magnitudes ($-1 > M_K > -6$) of Galactic globular clusters .  Their
relation is given in Equation \ref{eqn:rgb_slope}.

\begin{equation}
{\rm [Fe/H]} = -2.98 - 23.84 \times slope_{GB}
\label{eqn:rgb_slope}
\end{equation}

To apply this relation, we measure the RGB slope in the range $-3.25 >
M_K > -6$, iteratively throwing away $3 \sigma$ outliers; the faint
limit of $M_K=-3.25$ is where the measured LF begins to turn over due to
incompleteness (see \S \ref{sec:lfs}).  We find a slope of $-0.114 \pm
0.005$, which implies a mean metallicity of [Fe/H]$= -0.26 \pm 0.27$.
This is our estimate for the average stellar population in the central
regions of M33 ($3''< r \lesssim 12''$).  This determination is in
agreement with the estimate of [Fe/H]$\sim -0.5$ by \citet{OConnell1983}
using population spectral synthesis of the central $\sim 5''$ of M33.

However, as we will show in the next section, there is clearly a young
stellar component present in the region imaged.  Since the GB slope is
also sensitive to age \citep{Tiede1997}, in the sense that at a constant
metallicity the slope increases (gets less negative) with decreasing
age.  Thus if all the stars (old and young) have the same metallicity,
the GB slope will be slightly more positive than for a purely old
population, and the Kuchinski relation will give a metallicity which is
too low.  However, the younger stars are probably more metal rich than
the mean of the old population, and hence their GB slope will be closer
to that of the older, metal-poor population than if they were of the
same metallicity.  In short, our metallicity estimate is dependent on
both the metallicities and ages of all the stars in the population.

\subsection{Population Age} \label{sec:age}

\begin{figure}[ht]
\epsscale{0.6}
\plotone{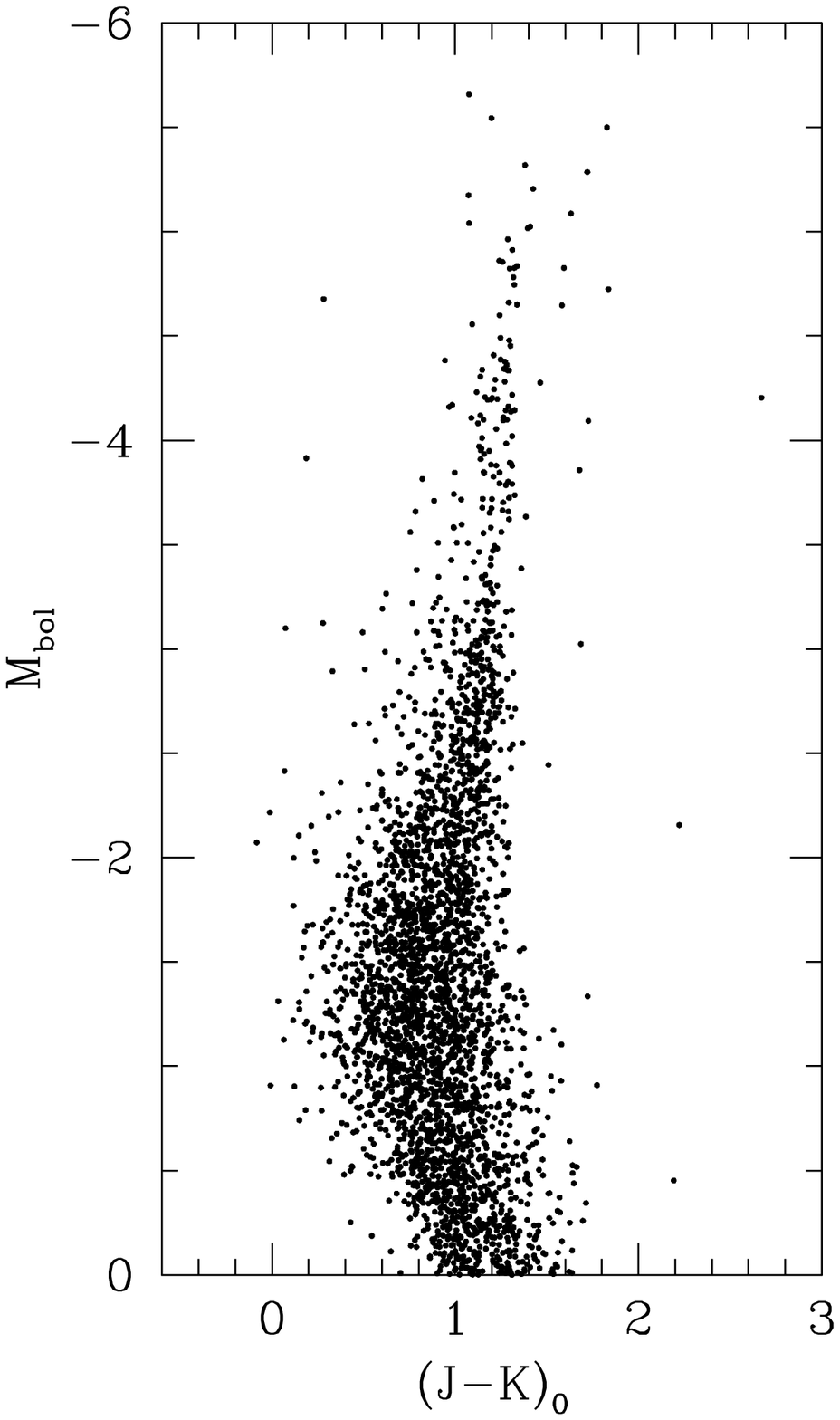} 
\figcaption{
Bolometric color magnitude diagram of the central regions of M33
($r>3''$).  Bolometric corrections are from \citet{Frogel1987}.  We
assume $(m-M)_0=24.64$, $A_K=0.03$, and $E(J-K)=0.05$.
\label{fig:cmd_mbol}}
\end{figure}

To estimate the age of the youngest stars in our field, we convert our
(CIT/CTIO) $K$-band measurements to bolometric luminosities using the
corrections of \citet{Frogel1987}.  These corrections were derived for
M-giants in Baade's Window and depend on the dereddened $(J-K)_0$ color.
The resulting bolometric CMD is shown in Figure \ref{fig:cmd_mbol}.

Assuming that the few bright stars at the tip of the AGB are members of
a young population, we can estimate the age of this population using the
relationship between AGB tip luminosity and age.  First used by
\citet{Mould1979, Mould1980}, this relation makes use of the
monotonically decreasing maximum luminosity of the AGB tip with
increasing age.  Due to the limited number of stars in our field, and
the short lifetime of stars on the AGB, this estimate is only an upper
limit to the age of the youngest stars.

\citet{Stephens2002} have recalculated the relation between the AGB tip 
luminosity and age using the ZVAR synthetic CMD code of
\citet{Bertelli1992}.  This program makes use of the stellar evolution
models of \citet{Bertelli1994}, and the mass-loss prescription of
\citet{Vassiliadis1993}.  Averaging three different models with
different metallicities and binary fractions, they derived Equation
\ref{eqn:age} for the ages of stars on the AGB tip at different 
luminosities.  Thus by measuring the bolometric luminosity of the AGB
tip, one can estimate the age of the youngest stars in the field.

\begin{eqnarray}
\log(Age)= & -0.91 - 7.962 M_{bol} - \nonumber \\
           & 1.7876 M_{bol}^2 - 0.12033 M_{bol}^3
\label{eqn:age}
\end{eqnarray}


The simplest estimate of the age comes from the luminosity of the
brightest star in the field.  This star has $M_{bol} = -5.658$, which,
with equation \ref{eqn:age}, gives an age of 0.51 Gyr.  

If we assume that the intermediate age stars on the AGB can be explained
by a single burst of star formation, we can apply the averaging
technique of \citet{Mould1982} to overcome small number statistics when
estimating the tip of the AGB.  Their method is to find the average
luminosity of stars on the AGB over the AGB tip of an old population
($M_{bol} \sim -4.5$).  By assuming that the distribution of stars along
the AGB is uniform, the peak bolometric luminosity for a fully populated
AGB should be twice the mean (plus $-4.5$).  Using this procedure on the
28 stars brighter than $M_{bol} = -4.5$ we estimate the AGB tip
luminosity to be $M_{bol} = -5.41 \pm 0.11$ for a fully populated AGB.
This luminosity implies an age of $0.79^{+0.18}_{-0.15}$ Gyr for this
intermediate age population.

However, looking at the bolometric CMD in Figure \ref{fig:cmd_mbol}, it
is fairly clear that the stars above $M_{bol} \sim -3.3$ are not
uniformly populated along its entire extent.  This is probably due to
multiple bursts of star formation contributing to the AGB.  A careful
inspection shows that the AGB is well represented from $-3.4 \gtrsim
M_{bol} \gtrsim -5$, with a bright tail of stars up to $M_{bol} \sim
-5.6$ (see also the luminosity function in Figure \ref{fig:lf_mbol}).
The predominant AGB is most likely due to a large burst of star
formation around 2 Gyr ago, while the less populated secondary AGB is
probably due to an even more recent burst of star formation $\sim 0.5$
Gyr ago.

A similar multi-generation model was proposed for the nucleus of M33 by
\citet{Gallagher1982} to explain its blue ($U-B$) and red ($B-V$) colors
and strong spectral absorption features such as H$\beta$ and
H$\gamma$. \citet{OConnell1983} also argued for multiple epochs of star
formation based on the integrated optical spectrum of the inner $10''$,
and concluded that $\sim 50$\% of the $V$-band light originates in stars
younger than 1 Gyr, with the youngest generation about $5 \times 10^6$
years old.  Ultraviolet observations agree, \citet{Ciani1984} used IUE
spectra together with $U,B,V$ photometry of the nucleus to model the
stellar population, and found the best-fit model required both a young
component with an age of $\sim 10^7$ years, and an old component with an
age of $\sim 10^{10}$ years.  While combining photometric and
spectroscopic observations, \citet{Gordon1999} found that the nucleus of
M33 is best fit by a $70-75$ Myr old single burst of star formation
enshrouded by a significant amount of dust.  Most recently,
\citet{Long2001} used stellar spectral synthesis to model HST STIS
spectroscopy of the nucleus and found that the best fit is with two
bursts of star formation 50 Myr and 1 Gyr ago.

%
%

We can reject the possibility of field star contamination as an
explanation of the bright stars observed in our field using the field
star density tabulated by \citet{Ratnatunga1985} which was calculated
from the Bahcall and Soneira Galaxy model. To estimate the number of
field stars brighter than $K=17$ ($M_K=-7.5$), we assume that potential
contaminants are field M-dwarfs, they could be as red as $(V-K) \sim 7$.
Thus they could be as faint as $V \sim 24$.  \citet{Ratnatunga1985}
predicts that the number of field stars with $V$ brighter than 25 is
$N(V<25)=3.124$ arcmin$^{-2}$.  Multiplying by the area of our field
0.116 arcmin$^2$ gives an upper limit of 0.36 stars brighter than $K=17$
in our field.  This means that there is a $<36\%$ chance of finding {\em
one} star in our field as bright as $K=17$, while we see 27.

\section{Luminosity Functions} \label{sec:lfs}

\begin{figure}[ht]
\epsscale{1}
\plotone{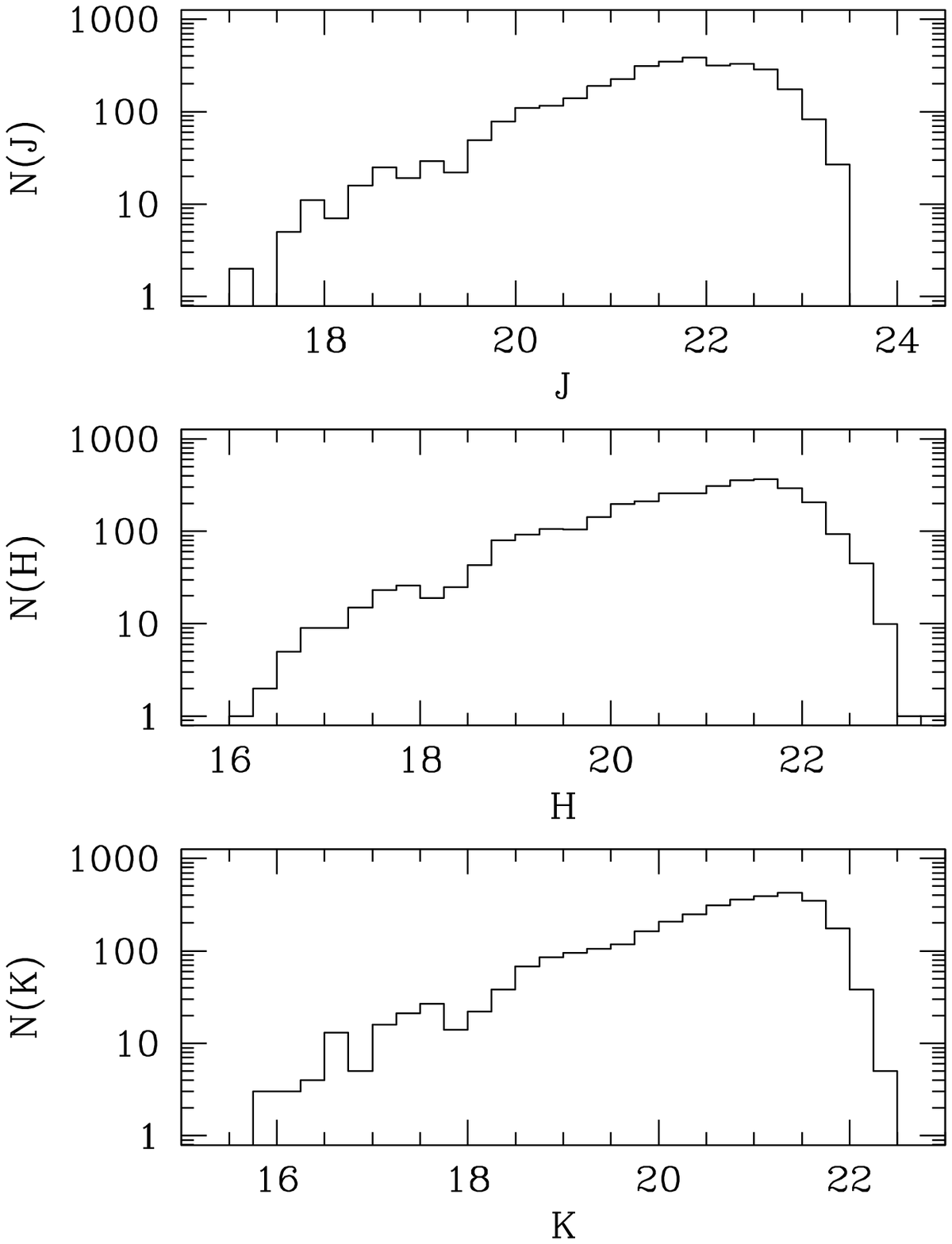} 
\figcaption{
The luminosity functions measured for $r>3''$ binned into 0.25 magnitude
bins.  The top panel is measured in $J$, the middle in $H$, and the
lower in $K$. 
\label{fig:lfs}}
\end{figure}

The luminosity functions (LFs) derived for our field are shown in Figure
\ref{fig:lfs} and listed in Table \ref{tab:lfs}.  These include only
objects measured farther than $3''$ from the center of M33.  This region
has an area of 416 arcsec$^2$.  The data have been binned into 0.25
magnitude bins, and the center of each bin is given in column 1.  These
LFs all show dips at the location of the tip of the RGB ($K \sim 18$).

\begin{deluxetable}{crrr}
\tablewidth{0pt}
\tablecaption{Luminosity Functions}
\tabletypesize{\footnotesize}
\tablehead{
\colhead{mag}	& 
\colhead{$N_J$}	&
\colhead{$N_H$}	&
\colhead{$N_K$}	}
\startdata
  15.875 &   0 &   0 &   3 \\ 
  16.125 &   0 &   1 &   3 \\ 
  16.375 &   0 &   2 &   4 \\ 
  16.625 &   0 &   5 &  13 \\ 
  16.875 &   0 &   9 &   5 \\ 
  17.125 &   2 &   9 &  16 \\ 
  17.375 &   0 &  15 &  21 \\ 
  17.625 &   5 &  23 &  27 \\ 
  17.875 &  11 &  26 &  14 \\ 
  18.125 &   7 &  19 &  22 \\ 
  18.375 &  16 &  25 &  38 \\ 
  18.625 &  25 &  43 &  68 \\ 
  18.875 &  19 &  80 &  85 \\ 
  19.125 &  29 &  92 &  95 \\ 
  19.375 &  22 & 106 & 105 \\ 
  19.625 &  49 & 105 & 118 \\ 
  19.875 &  78 & 142 & 162 \\ 
  20.125 & 110 & 198 & 208 \\ 
  20.375 & 116 & 213 & 249 \\ 
  20.625 & 139 & 256 & 310 \\ 
  20.875 & 190 & 258 & 359 \\ 
  21.125 & 226 & 310 & 388 \\ 
  21.375 & 310 & 357 & 427 \\ 
  21.625 & 350 & 364 & 349 \\ 
  21.875 & 385 & 293 & 176 \\ 
  22.125 & 318 & 207 &  38 \\ 
  22.375 & 330 &  94 &   5 \\ 
  22.625 & 286 &  45 &   0 \\ 
  22.875 & 175 &  10 &   0 \\ 
  23.125 &  83 &   1 &   0 \\ 
  23.375 &  27 &   0 &   0 \\ 
\enddata
\label{tab:lfs}
\end{deluxetable}

\begin{figure}[ht]
\epsscale{1}
\plotone{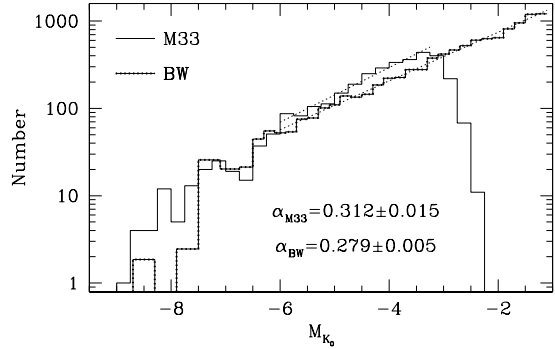} 
\figcaption{
Comparison between luminosity functions measured in M33 (solid line),
and the bulge of our Galaxy viewed through Baade's Window (beaded
line).  The M33 LF is constructed by placing the stars measured outside
a $3''$ radius into 0.25 magnitude wide bins.  The BW LF is the
combination of the bright end of \citet{Frogel1987}, and the faint end
of \citet{DePoy1993}, and has been scaled to match the number of M33
counts in the magnitude range $-7.5 < M_K < -6$.  This plot assumes that
$(m-M)_{M33}=24.64$ and $(m-M)_{BW}=14.5$
\label{fig:lfcompbw}}
\end{figure}

We show the absolute $K$-band luminosity function in Figure
\ref{fig:lfcompbw}, assuming a distance modulus to M33 of 24.64.
For comparison we have also plotted a composite Galactic Bulge
luminosity function constructed from \citet{Frogel1987} and
\citet{DePoy1993} measured in Baade's Window (BW).

There are two very obvious differences between these two LFs: their
bright end extents and their slopes.  The M33 LF extends over a
magnitude brighter than what is observed in BW.  These bright stars, as
discussed in Section \ref{sec:age}, are a result of a young population
of stars in the central regions of M33.

The slopes of the M33 and Baade's Window RGB LFs are also significantly
different.  We fit each with a single power law, M33 over $-6 < M_K <
-3.25$, and BW over $-6 < M_K < -1.1$.  In M33 we measure a LF slope of
$0.312 \pm 0.015$, while in BW we find $0.279 \pm 0.005$.  One possible
reason for this difference is the difference in mean ages of the stars
in these two regions.  M33 contains a mix of young and old stars (e.g.
Section \ref{sec:age} and the optical-IR CMD in Figure
\ref{fig:cmd_vk}), while the Galactic bulge is predominantly an old
population.  However, \citet{Davidge2000b} has also suggested that the
slope of the LF may vary throughout the Galactic bulge.  Looking at
dereddened LFs of 17 bulge fields, he fits a power law between
$M_K=-0.5$ and $+1.0$ and finds a range in LF slope of $0.165 \pm 0.064
< \alpha < 0.672 \pm 0.225$, with a mean value, obtained by coadding the
LFs, of $0.335 \pm 0.018$.  This mean LF slope obtained by Davidge
agrees with what we have measured in the central region of M33, however
involves many uncertainties, such as the reddening and the small
luminosity range over which the LFs were fit.  If indeed galaxy LFs are
variable, as Davidge's data suggests, deeper high-resolution imaging of
M33 may prove one of the most robust ways to verify this effect.


\begin{figure}[ht]
\epsscale{1}
\plotone{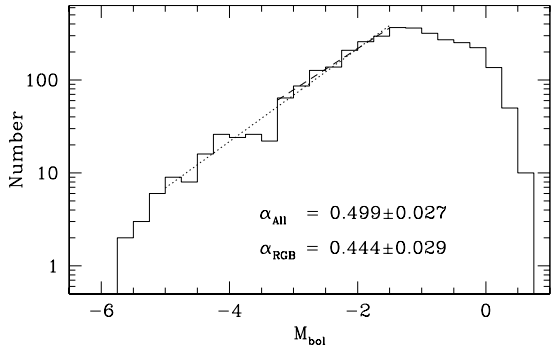} 
\figcaption{
Bolometric luminosity function for all objects measured farther than
$3''$ from the nucleus of M33.  Bolometric corrections are from
\citet{Frogel1987}, where we have assumed that $(m-M)_0=24.64$,
$A_K=0.03$, and $E(J-K)=0.05$.  A fit to the RGB (dashed line; $-3.25 <
M_{bol} < -1.5$) gives a slope of $0.444 \pm 0.029$, while fitting the
entire LF (dotted line; $-5 < M_{bol} < -1.5$) yields a slope of $0.499
\pm 0.027$.
\label{fig:lf_mbol}}
\end{figure}

Finally we present the bolometric luminosity function derived from our
observations in Figure \ref{fig:lf_mbol}.  The data cover 416 square
arcseconds, avoiding the central $3''$.  We have used the bolometric
corrections from \citet{Frogel1987} which are based on each star's
$(J-K)$ color.  Here the dip in the LF at the RGB-AGB transition
$(M_{bol} \sim -3.5)$ is especially apparent.  The termination of the LF
at $M_{bol} \sim -5.5$ is in agreement with the youngest stars on the AGB
having an age $\sim 0.5$ Gyr (see Section \ref{sec:age}).

Fitting a power-law to the LF of the RGB ($-3.25 < M_{bol} < -1.5$) we
find a slope of $\alpha_{RGB} = 0.444 \pm 0.029$.  If instead we fit the
entire luminosity function from the bright end dropoff at $M_{bol} = -5$
to the completeness limit at $M_{bol} = -1.5$, we find a slope
$\alpha_{All} = 0.499 \pm 0.027$.  \citet{Davidge2000a} measured the
slope between $-5 < M_{bol} < -3.5$, a much smaller range than ours, and
found a slope of $0.528 \pm 0.036$, in good agreement with our
determination.

\section{Comparison with Previous Data} \label{sec:comparison}

\subsection{\citet{Davidge2000a}} \label{sec:davidge}

Recently \citet{Davidge2000a} has made AO observations of the nuclear
regions of M33 with the 3.6m Canada-France-Hawaii Telescope (CFHT).  His
$\sim 35''$ field had a total of 20 minutes exposure time per $JHK$
filter, and his images have a FWHM of $0.34''$, about twice the size of
the Gemini PSF, which ranged from $0.12''$ to $0.20''$.

\begin{figure}[ht]
\epsscale{1}
\plotone{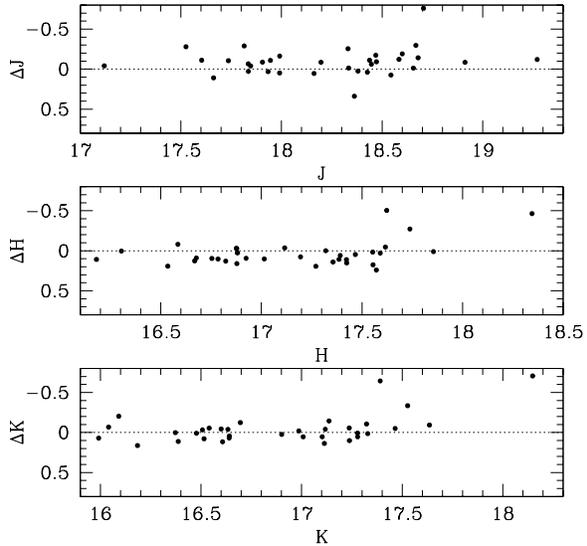} 
\figcaption{
Comparison between measurements made by \citet{Davidge2000a} with the
3.6m CFHT, and the present Gemini-North observations.  The mean
difference, excluding three sigma outliers is $\Delta J = -0.08 \pm
0.02$ and $\Delta H = 0.08 \pm 0.01$.  The $K$-band difference is by
definition zero ($\pm 0.02$) since it was used for calibration.
\label{fig:dm_compare_davidge}}
\end{figure}

Using his published list of photometry of stars with $K \le 17$, we have
matched up observations of 34 stars.  The difference between these
measurements is illustrated in Figure \ref{fig:dm_compare_davidge}.
Several of the fainter stars are obviously blended in Davidge's images,
however the agreement with the bulk of the sample is quite good.
Throwing out three sigma outliers, the average difference between the
samples (Davidge -- Gemini) is $\Delta J = -0.08 \pm 0.02$, $\Delta H =
0.08 \pm 0.01$, and $\Delta K = 0.00 \pm 0.02$.  The perfect agreement
in the $K$-band is of course because we used the same sample of good
measurements to determine our $K$-band photometric zero point.

\begin{figure}[ht]
\epsscale{1}
\plotone{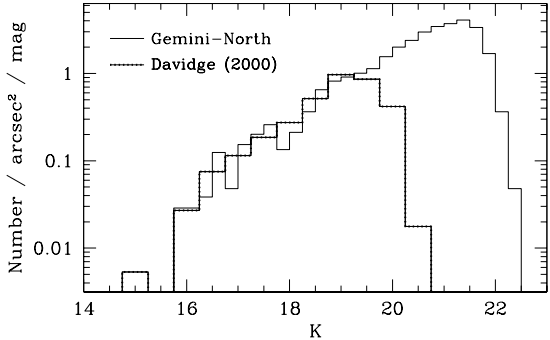} 
\figcaption{
Comparison between $K$-band luminosity functions measured in the central
regions of M33 with Gemini-North (solid line) and by \citet{Davidge2000a}
(beaded line).  Our Gemini data only includes objects measured farther
than $3''$ from the nucleus.  Both LFs have been normalized to show the
number of stars per square arcsecond per magnitude.  The brightest bin
($K=15$) of Davidge's LF contains only one star with $K=15.03$.
\label{fig:lf_compare_davidge}}
\end{figure}

The comparison between our $K$-band LFs is shown in Figure
\ref{fig:lf_compare_davidge}.  The two LFs agree very well, both in 
slope and normalization.  The most obvious difference is that our LF
extends to $K \gtrsim 21.5$, while the Davidge LF rolls over at $K\sim
19$ due to incompleteness.  The Davidge LF also has one star $\sim 1$
magnitude brighter than any star in our field, however this is most
likely a result of our smaller area not completely sampling the
brightest star population.  Another difference is that while we see a
small dip in the LF due to the tip of the RGB at $K \sim 18$, this
feature is not visible in the Davidge LF, probably because it is smeared
out due to his reduced photometric accuracy at this level.

\subsection{\citet{Mighell2002}} \label{sec:mighell}

\citet{Mighell2002} have observed the nuclear region of M33 using the
Wide Field Planetary Camera 2 (WFPC2) on the Hubble Space Telescope
(HST).  They have obtained images through the F555W ($\sim V$) and F814
($\sim I$) filters.  Their optical CMD reveals a bright blue main
sequence, a red supergiant plume, a very red AGB, and a wide RGB.  They
interpret this complex CMD as indicative of a mix of young ($<100$
Myr), intermediate age ($\sim 1$ Gyr), and old stellar populations ($>10$
Gyr).

\begin{figure*}[ht]
\epsscale{2}
\plotone{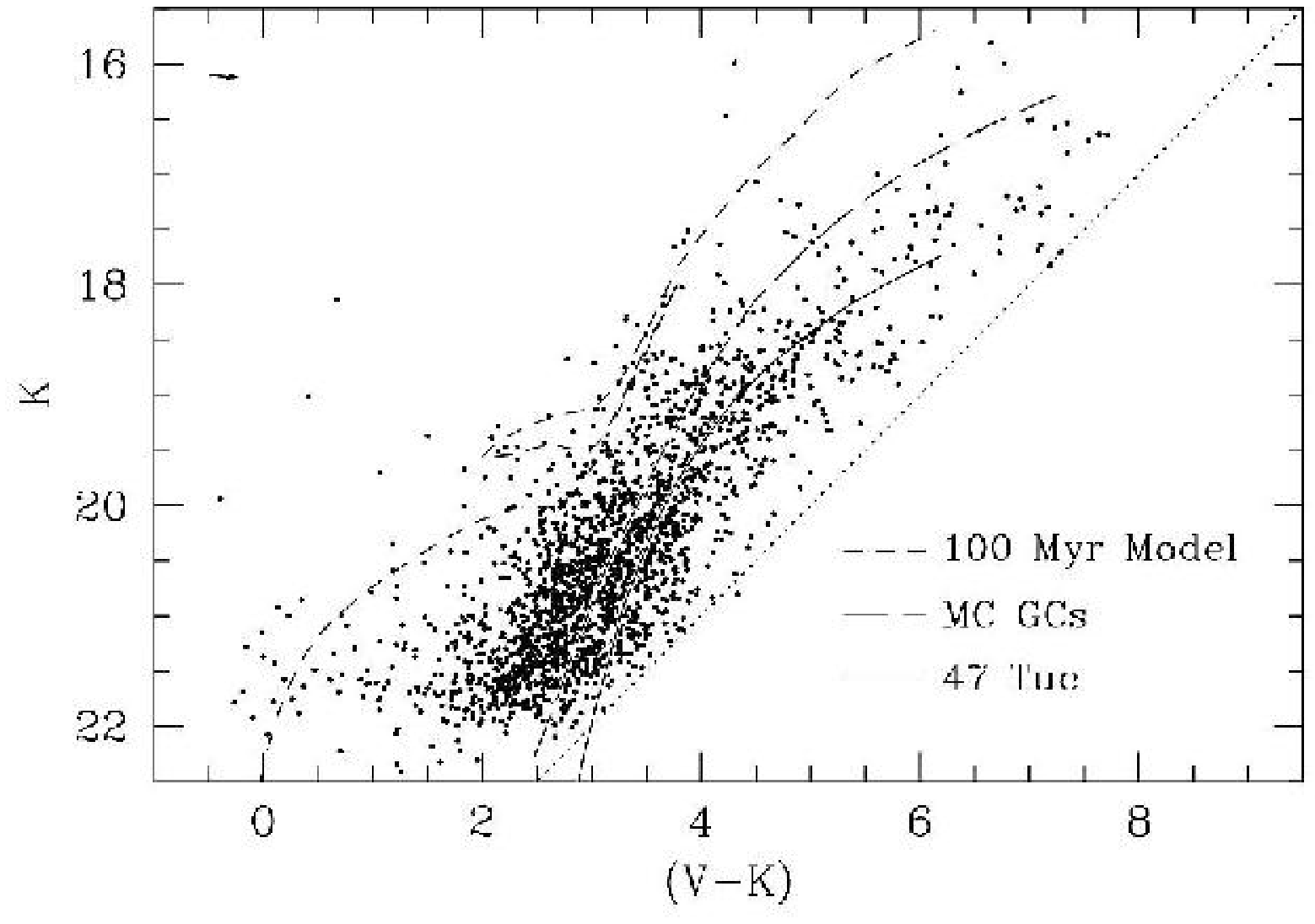} 
\figcaption{
Combined optical \citep[WFPC2,][]{Mighell2002} and infrared
(Gemini-North) color-magnitude diagram.  For comparison we have
overplotted the GB locus of the Galactic globular cluster 47 Tuc
([Fe/H]$=-0.76$) from \citet{Ferraro2000}, the mean ridge line of 12
intermediate age Magellanic Cloud clusters from \citet{Ferraro1995}, and
a 100 Myr solar metallicity isochrone from \citet{Girardi2002}.  The
overplotted lines have been reddened to match the assumed reddening in
M33, which is illustrated by the small arrow in the upper left corner.
The limiting magnitude of the optical photometry at $V=25$ is marked by
the dotted line.
\label{fig:cmd_vk}}
\end{figure*}

We have matched up their optical WFPC2 observations with our infrared
Gemini measurements using {\sc daomaster} \citep[part of the {\sc
daophot} package,][]{Stetson1987} and show the resulting $K-(V-K)$ CMD
in Figure \ref{fig:cmd_vk}.  For comparison we have overplotted the GB
of the old Galactic globular cluster 47 Tuc ([Fe/H] $=-0.76$) from
\citet{Ferraro2000}.  This cluster GB exemplifies the region of this CMD
which will be occupied by an old intermediate-metallicity stellar
population.  The color is fairly red, and the GB tip is well below
$K=17.5$.  We have also overplotted the mean ridge line of 12
intermediate age Magellanic Cloud clusters from \citet{Ferraro1995}.
The component clusters have an average SWB-type of {\sc IV}
\citep{Searle1980}, an average $s$ value of 36 \citep{Elson1985}, and a
mean metallicity of [Fe/H]$=-1.56$ based on the slope and placement of
their composite $(V-K)$ GB.  This GB exemplifies the location of an
intermediate age metal poor stellar population, with the tip of the GB
extending over a magnitude brighter than that of 47 Tuc.  Lastly, we
overplot a 100 Myr, $\sim$ solar metallicity ($Z=0.019$, $Y=0.273$)
isochrone from \citet{Girardi2002}, which uses a detailed TP-AGB
treatment.  This isochrone extends even bluer and brighter than the
other two RGBs, and can explain the large number of very blue stars with
$(V-K)<2$.

Figure \ref{fig:cmd_vk} makes it clear that, as concluded by
\citet{Mighell2002}, there is a lot going on in the central regions of
M33.  There appear to be young blue main sequence stars and red
supergiants, an intermediate age GB similar to the MC clusters, and an
evolved GB like that of the Galactic cluster 47 Tuc.  For a discussion
of previous work on this topic see the end of Section \ref{sec:age}.
Unfortunately, degeneracies between age and metallicity make it very
difficult to quantitatively describe the mix of stellar populations
based on our photometric dataset alone. \citep{OConnell1982,
Worthey1994}

\section {Radial Variations in M33 Populations} \label{sec:radial_variations}

In Section \ref{sec:decomposition} we showed that the best 3-component
fit to the surface brightness profile includes spheroid and disk
components whose relative contributions vary significantly across our
field.  If this model is correct, and these two populations are
sufficiently different, we may be able to detect this variation by
looking for a radial dependence of the slope and extent of the
luminosity function, and in the morphology of the color-magnitude
diagram.

\begin{figure}[ht]
\epsscale{1}
\plotone{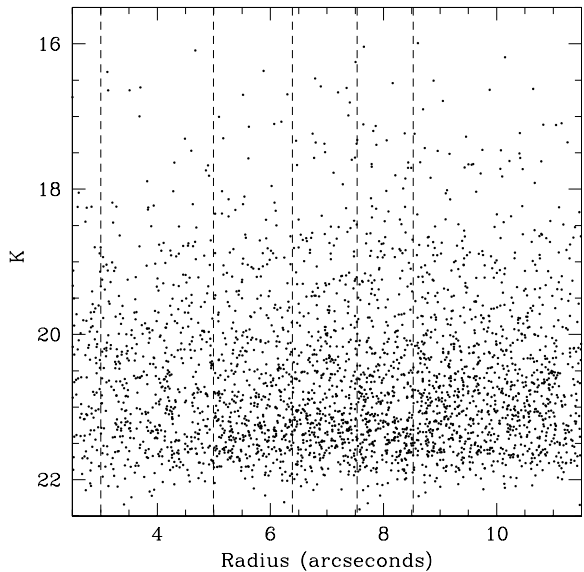} 
\figcaption{
The radial distribution of stars measured in the central region of M33.
The dashed lines at $r=$ 3, 5, 6.4, 7.5, and 8.5 arcseconds indicate the
boundaries of the regions used in the analysis of Section
\ref{sec:radial_variations}.
\label{fig:stellar_dist}}
\end{figure}

To test this hypothesis, we have divided our field into four equal-area
rings around the center of M33.  The results are summarized in Table
\ref{tab:radial_variations} which lists the properties measured in each
ring as well as in the entire field.  The first two columns of Table
\ref{tab:radial_variations} give the limits of each ring, chosen so that
each has an area of 50 arcseconds$^2$ (also illustrated in Figure
\ref{fig:stellar_dist}).  The third column ($N$) lists the number of
stars measured in each ring.  The fourth and fifth columns give the
giant branch slope ($m_{GB}$) and the luminosity function power-law
slope ($\alpha_{LF}$), both measured from $-3.25 > M_{K_0} > -6$.  The
last column gives the brightest star measured in each annulus as an
estimate of the tip of the AGB.  The results for the entire frame are
listed in the last row of the table, and marked $r>3''$.

\begin{deluxetable}{cccccc}
\tablewidth{0pt}
\tablecaption{Variations with Radius}
\tabletypesize{\footnotesize}
\tablehead{
\colhead{$R_1$\tablenotemark{a}}	&
\colhead{$R_2$\tablenotemark{a}}	&
\colhead{$N$}				&
\colhead{GB slope\tablenotemark{b}}	&
\colhead{$\alpha_{LF}$\tablenotemark{b}}&
\colhead{$M_{bol}$(AGBT)}		}
\startdata
3.0 & 5.0 & 484 & $-0.118 \pm 0.012$ & $0.260 \pm 0.029$ & $-5.32$ \\
5.0 & 6.4 & 467 & $-0.082 \pm 0.013$ & $0.333 \pm 0.036$ & $-5.02$ \\
6.4 & 7.5 & 444 & $-0.115 \pm 0.012$ & $0.338 \pm 0.055$ & $-5.17$ \\
7.5 & 8.5 & 424 & $-0.094 \pm 0.012$ & $0.316 \pm 0.039$ & $-5.29$ \\
\multicolumn{2}{c}{
$r>3''$} & 3308 & $-0.114 \pm 0.005$ & $0.312 \pm 0.015$ & $-5.66$ \\
\enddata
\tablenotetext{a}{Radii in arcseconds.}
\tablenotetext{b}{Measured from $-3.25 > M_K > -6.0$}
\label{tab:radial_variations}
\end{deluxetable}

\subsection{Variations in the Stellar Distribution}

In this section we analyze both the radial distribution of stars, as
indicated by the Kolmogorov-Smirnov statistic (KS-test), and the
distribution of stellar luminosities as measured by fitting power-law
luminosity functions as well as the KS-test.  The radial distribution of
measured $K$-band magnitudes are shown in Figure \ref{fig:stellar_dist},
which illustrates the continuum of stellar luminosities over the field.

First we perform the KS-test on the radial positions of stars plotted in
the optical-IR CMD (Figure \ref{fig:cmd_vk}).  Here we have divided the
stars into two groups: blueward of the intermediate age Magellanic Cloud
cluster locus, which we call ``young'', and redward of the MC cluster
locus, which we call ``old''.  For both groups we consider stars only
brighter than $K=21$ to minimize the effects of incompleteness.  When
considering all stars from $3''$ to $14''$, the KS-test shows a marginal
difference ($P=0.01$) between the radial distribution of the ``young''
and ``old'' stars, however this difference arises in the very central
regions, and is most likely due to incompleteness in this more crowded
region.  When we limit the test to stars outside $4''$, the radial
distributions of the two groups become nearly indistinguishable
($P=0.08$).  Thus based on the stars from $4''$ to $14''$ in the
optical-IR CMD, we would conclude that the radial distributions of
``young'' and ``old'' stars are identical.

If instead, we consider our whole sample of infrared measurements, and
are more selective in our choice of ``young'' stars, namely bluer than
$(J-K)=0.6$ and brighter than $K=21.5$, these stars appear to lie at
larger radii than the remainder of the population.  Performing a KS-test
on this small subsection of the CMD, from $4''$ to $14''$, gives a very
low probability ($P=2.7E-6$) that these stars have the same radial
distribution as the other stars.  However the observed difference arises
at primarily large radii ($r \gtrsim 9''$).  If the KS-test is run from
$4''$ to $10''$ the significance completely disappears.  We mention this
result as an aside because it is only evident in the outermost regions
of the field, where the completeness due to dithering is difficult to
calculate and the probability for systematic effects caused by the large
distance from the wavefront reference source (nucleus) is highest.

\citet{Mighell1995} did find significantly different radial distributions 
of young and old stars in the central $\sim 70''$ of M33 based on
optical HST-PC observations.  Specifically they found that the younger
Pop I stars preferentially lie farther from the nucleus than the more
centrally concentrated older Pop II stars.  

Next we look at the luminosity functions measured in each of the four
rings defined in Table \ref{tab:radial_variations} and illustrated in
Figure \ref{fig:stellar_dist}.  We fit a power-law to the RGB from $-6 <
M_{K_0} < -3.25$ and find that there is no significant change across the
field, and that the slopes determined are all consistent with that
measured for the entire field: $\alpha = 0.312 \pm 0.015$.  A more
rigorous analysis using the KS-test verifies this general result.  These
tests show that based on the distribution of stellar luminosities in
each of the four annuli, as well as on the radial distribution of stars
of different luminosities (binned into 1-magnitude bins), all stars are
consistent with being drawn from the same population.

\citet{McLean1996} obtained similar results for the inner disk of M33.
Comparing AGB stars brighter than $K \sim 18$ in two regions, $45'' < r
< 1.5'$ and $1.5' < r \lesssim 4'$, they found no significant difference
between the luminosity functions except for the presence of luminous
supergiants in the outer region.

\begin{figure}[ht]
\epsscale{1}
\plotone{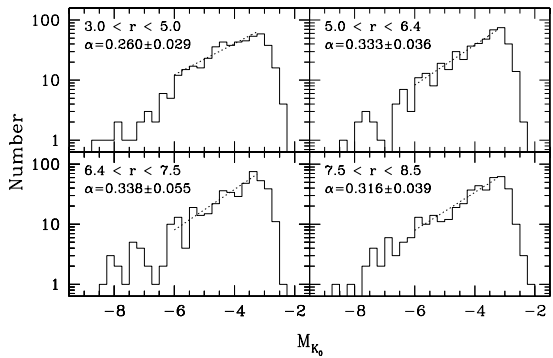} 
\figcaption{
Luminosity functions for four equal area rings around M33.  A power-law
has been fitted to each in the range $-6 < M_{K0} < -3.25$ and is
overplotted with a dotted line.  The slopes are given in the upper left
corner of each panel.
\label{fig:lf_4radii}}
\end{figure}

\subsection{Variations in the CMD}

\begin{figure}[ht]
\epsscale{1}
\plotone{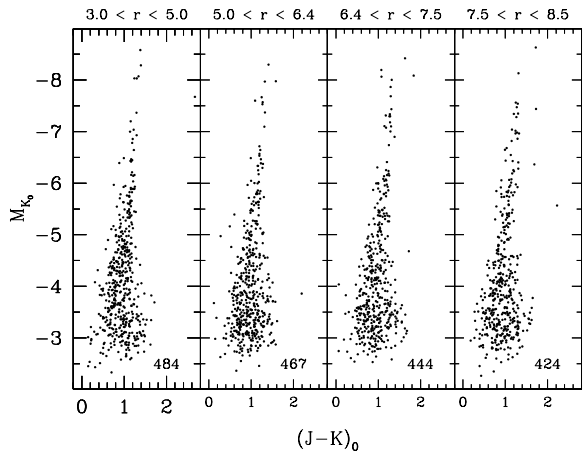} 
\figcaption{
Color-magnitude diagrams for rings of equal area around M33.  The number
of stars in each ring is given in the lower right corner of each panel.
\label{fig:cmd_4radii}}
\end{figure}

The CMDs of each of the four rings are displayed in Figure
\ref{fig:cmd_4radii}.  The radial limits of each are at the top of
each CMD, and the number of objects found in each ring is printed in the
bottom right corner.  Using the same technique which we used to estimate
the mean metallicity of the stellar population in Section \ref{sec:cmds}, 
we now measure the slope of the GB in each ring.  As before, the
iterative linear fit is only to data with $-3.25 > M_{K0} > -6$, and
ignores three sigma outliers.

The results of the GB fitting are listed in column 4 of Table
\ref{tab:radial_variations}.  We find that based on the slope of
the giant branch, the mean metallicity of each annulus is consistent
with that of the entire field.  This is similar to what has been
observed in the inner Bulge of our Milky Way ($R<560$ pc), which shows
no evidence for a gradient along the major or minor axes
\citep{Ramirez2000}.

Although we have not detected a gradient in the stellar properties in
the inner $\sim 10''$, we have nonetheless helped fill in the gap
between the metallicity gradient observed in the disk of M33
\citep{Henry1995, Kwitter1981, Searle1971}, and low abundance
measurements for the nuclear region, most recently [Fe/H]$= -1.2 \pm
0.5$ ($0.5'' > r > 1''$) based on the CO index by \citet{Davidge2000a}.



\section{Luminous Stars and Blending in M33} \label{sec:blending}

\begin{figure}[ht]
\epsscale{1}
\plotone{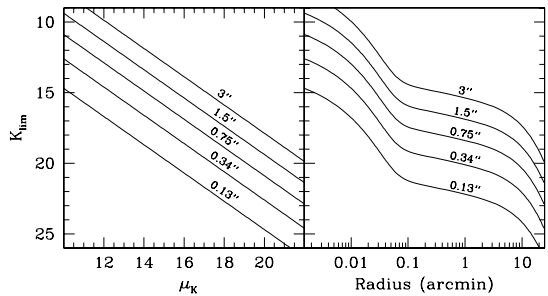} 
\figcaption{
Limits blending places on stellar photometry in M33, as determined by
the general rule that meaningful photometry can only be obtained for
stars brighter than the luminosity sampled by a single resolution
element \citep{Renzini1998}.  The left panel gives the $K$-band
magnitude ($K_{lim}$) of the faintest stars which can be accurately
measured (against blending) as a function of the background surface
brightness for five different imaging resolutions.  The right panel uses
the surface brightness profile of M33 to convert surface brightness to
radial distance from the center of M33.  Our resolution is $0.13''$, and
\citet{Davidge2000a} had a resolution of $0.34''$.
\label{fig:blendinglimits}}
\end{figure}

As \citet{Renzini1998} has pointed out, meaningful photometry can only
be obtained for stars brighter than the luminosity contained in each
resolution element.  We have calculated the enclosed luminosity in M33
for five different imaging resolutions ($0.13''$, $0.34''$, $0.75''$,
$1.5''$, and $3''$), assuming that the size of the resolution element is
defined by the FWHM.  Figure \ref{fig:blendinglimits} shows the
corresponding limiting $K$-band magnitude as a function of surface
brightness.  This plot shows that at $\mu_K \sim 16.8$, the surface
brightness over most of our field, the faintest stars which we can
accurately measure with our $\sim 0.13''$ resolution have $K \sim 21.5$.
This is in very good agreement with the observed limit of our photometry
(e.g. Figure \ref{fig:cmd}).

Even more useful perhaps, is a plot of the limiting magnitude due to
blending as a function of spatial position.  For symmetric surface
brightness distributions, such as exist around M33, the conversion is
relatively simple.  Using our three component surface brightness model
(Section \ref{sec:decomposition}), we have transformed from generic
surface brightness to the radial distance from the center of M33
measured in arcminutes.  The resulting relations for the same five
different resolutions are shown on the right side of Figure
\ref{fig:blendinglimits}.  Thus for our data, with a resolution of $\sim
0.13''$, the faintest stars which we can accurately measure (against
blending) at $6''$ arcseconds from the center of M33 have $K \sim 21.3$.
If we try to measure stars at a distance of $3''$, blending will limit
our measurements to only stars brighter than $K \sim 20.3$, a full
magnitude brighter than we could accurately measure at $6''$.

Another simple calculation advocated by \citet{Renzini1998} is the
estimation of the number of stars in each evolutionary stage per
resolution element.  This allows one to estimate the severity of
blending in any observation.  This calculation has several parameters
which have a weak dependence on the age, metallicity, and IMF of the
stellar population.  The ratio of total ($L_T$) to $K$-band luminosity
($L_K$), and the specific evolutionary flux, $B(t)$, are two such
parameters.  For our calculation we used $L_T / L_K = 0.36$, and $B(t)=
2.2 \times 10^{-11}$ stars yr$^{-1}$ \lsun$^{-1}$, numbers suitable for
a 15 Gyr old, solar-metallicity population.

\begin{figure}[ht]
\epsscale{1}
\plotone{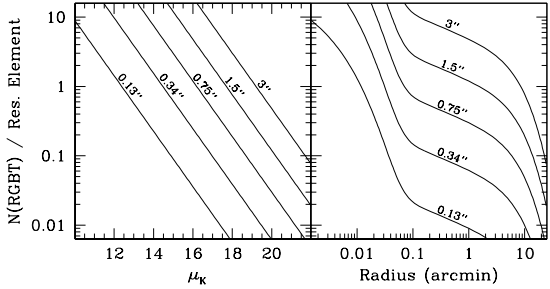} 
\figcaption{
The number of M33 RGB stars within one magnitude of the RGB tip per
resolution element, based on the formulae of \citet{Renzini1998}.  The
probability of a blend of two RGBT stars scales as the square of
N(RGBT).  The left panel shows N(RGBT) as a function of the background
surface brightness for five different imaging resolutions.  The right
panel uses the surface brightness profile of M33 to convert surface
brightness to radial distance from the center of M33.
\label{fig:nrgbt_reselement}}
\end{figure}

The results of this calculation for stars within one magnitude of the
RGB tip (RGBT) are displayed in Figure \ref{fig:nrgbt_reselement}.  In
the left panel we show the number of RGBT stars per resolution element
as a function of surface brightness for five different imaging
resolutions.

The number of blends of two RGBT stars on a frame can be estimated as
the square of the number per resolution element (for $N<1$) multiplied
by the number of resolution elements in the frame \citep{Renzini1998}.
If the number of RGBT stars per resolution element is greater than one,
it should be clear that photometry of stars at or below the level of the
RGBT is impossible.  At this point only stars several magnitudes
brighter than the RGBT can be measured with any accuracy, but then one
must face the question of whether the objects measured are real or just
blends of many stars, each of which can be as bright as the tip of the
RGB.

\citet{Stephens2001} have performed simulations of the blending in 
their NICMOS observations of globular clusters in M31.  They show that
severe blending can easily create objects which are several magnitudes
brighter than any star in the parent population.  Thus one must be very
careful when interpreting bright objects measured in a very crowded
fields, i.e. where N(RGBT) per resolution element is greater than one.

It might be illuminating to examine a few previous studies of the
central regions of M33 keeping Figures \ref{fig:blendinglimits} and
\ref{fig:nrgbt_reselement} in mind.  The most critical, at least in
terms of our photometric calibration, are the observations of
\citet{Davidge2000a} discussed in Section \ref{sec:data_reduction}.
Plotting his $0.34''$ resolution on both Figures, we see that his
observations are not significantly affected by blending.  The right side
of Figure \ref{fig:nrgbt_reselement} shows that Davidge stays below 1
RGBT star per resolution element until only $2''$ from the nucleus,
which he (for the most part) steers clear of.  Assuming the average
distance of his field from the nucleus is $\sim 9''$, Figure
\ref{fig:blendinglimits} shows that the limit of accurate photometry
(against blending) is $K \sim 19.4$.  Since his estimated completeness
limit is $K=19$, his photometry is therefore limited by photon noise and
not blending.

Checking the observations of \citet{McLean1996}, we see that their
infrared $JHK$ observations of the central $7.6'$ of M33 had $1.5''$
seeing.  They divided their observations into a ``Central Core Region''
going from $45''$ to $1.5'$, and an ``Inner Disk'' region, spanning from
a radius of $1.5'$ to $\sim 4'$.  Looking at the $1.5''$ resolution
curve on Figure \ref{fig:blendinglimits}, we find that at $1.5'$, their
photometric limit imposed by blending is $K \sim 17$, and Figure
\ref{fig:nrgbt_reselement} shows that at this distance there is $\sim 1$
RGBT star per resolution element.  Since the number of detected stars in
their core region abates at $K \sim 17.5$, they are not trying to go
much deeper than blending allows.  However, the high density of stars as
bright as the RGB tip, and the fact that out to $15''$ we only find
stars as bright as $K \sim 16$, suggests that some of the brighter ($K
\sim 14.5$), bluer [$(J-K)<1.5$] objects they detected in the central
region may be blends.

The observations of \citet{Minniti1993, Minniti1994} are in a similar
situation in terms of blending.  With $1.5''$ resolution they measured
stars as bright as $K \sim 14$ in the inner $2'$ of M33.  However, as
Figure \ref{fig:nrgbt_reselement} shows, at $2'$ with $1.5''$
resolution, there are $\sim 0.9$ stars within one magnitude of the tip
of the RGB per resolution element.  This certainly does not guarantee
that these bright objects are blends, but merely shows that the
potential for blending is high.


\section{Conclusions} \label{sec:conclusions}

We have used Gemini-North to study the stellar populations in the
central regions of M33.  The surface brightness profiles from $0.1''$ to
$18'$, formed from the combination of our data and those of
\citet{Regan1994}, show that the data need to be modeled using a
three-component, core + spheroid + disk model.  The best-fit parameters
are listed in Table \ref{tab:decomposition}.

These high-resolution observations allow us to accurately measure
individual stars to $K \sim 21$.  Artificial star tests (\S
\ref{sec:artstartests}) show that our completeness is relatively
uniform across the field (50\% at $K=21$), although within $3''$ from
the nucleus the completeness is dramatically lower due to severe
crowding.  Artificial fields are used to understand the observational
effects associated with adaptive optics measurements in crowded fields.

Based on the slope of the giant branch in the infrared color magnitude
diagram, we estimate the mean metallicity to be $-0.26 \pm 0.27$ (\S
\ref{sec:metallicity}).  Using the bolometric luminosities and density
of stars on the AGB, we hypothesize two bursts of star formation; at
$\sim 0.5$ and $\sim 2$ Gyr ago (\S \ref{sec:age}).  We note however,
that this component of young stars may have influenced our metallicity
estimate due to the sensitivity of the GB slope on age.

The stellar luminosity function in M33 is shown to be significantly
different from that measured in the Galactic Bulge as viewed through
Baade's Window (\S \ref{sec:lfs}).  The difference in their maximum
luminosities is due to differences in ages of the two regions.  We
speculate that this is also the origin of their different slopes as
well.

In section \ref{sec:comparison} we compare our data with previous
observations.  Recent work by \citet{Davidge2000a} is in good agreement
with our data, although our observations go $>2$ magnitudes deeper (\S
\ref{sec:davidge}).  We also combine our data with optical HST-WFPC2
measurements \citep{Mighell2002}, and present the optical-IR CMD in \S
\ref{sec:mighell}.  This CMD clearly shows that the central regions of
M33 are composed of young, intermediate, and old aged stellar
populations.

Dividing the inner $\sim 8.5''$ ($\sim 35$ pc) into equal area rings
around the nucleus, we look for radial variations in the stellar
properties (\S \ref{sec:radial_variations}).  However, based on the
distribution of stellar luminosities and the morphology of the CMD, we
find that all stars are consistent with being drawn from a single
population.

In the last section (\S \ref{sec:blending}) we perform calculations to
estimate the severity of blending at various imaging resolutions and
locations in M33.  Using the formulations of \citet{Renzini1998} and our
composite surface brightness profile, these calculations call into
question some previous claims of very luminous stars in the central
regions of M33.

\acknowledgements

Support for this work was provided by a Presidential Fellowship from the
Graduate School at the Ohio State University, and by a Princeton-
Catolica Prize Fellowship, both awarded to AWS.

This paper was based on observations obtained at the Gemini Observatory,
which is operated by the Association of Universities for Research in
Astronomy, Inc., under a cooperative agreement with the NSF on behalf of
the Gemini partnership: the National Science Foundation (United States),
the Particle Physics and Astronomy Research Council (United Kingdom),
the National Research Council (Canada), CONICYT (Chile), the Australian
Research Council (Australia), CNPq (Brazil) and CONICET (Argentina).
Observations used the Adaptive Optics System Hokupa'a/QUIRC, developed
and operated by the University of Hawaii Adaptive Optics Group, with
support from the National Science Foundation.

Many thanks to the Gemini queue observers, Peter Stetson for supplying
and helping us with his photometry package, ALLFRAME, and to Ken Mighell
for providing us with his HST-WFPC2 images and photometry lists before
publication.  Thanks to Michael Regan for helping with his surface
brightness decomposition.  Valuable comments from Ken Mighell, Darren
DePoy, Dante Minniti, and an anonymous referee were also greatly
appreciated.

\end{document}